\documentclass[12pt,a4paper]{article}
\usepackage{amsmath,amssymb,amsthm} \numberwithin{equation}{section}

\theoremstyle{definition}

\theoremstyle{remark}

\newcommand{\beq}{\begin{eqnarray}}
\newcommand{\eeq}{\end{eqnarray}}
\newcommand{\beqnn}{\begin{eqnarray*}}
\newcommand{\eeqnn}{\end{eqnarray*}}
\newcommand{\rd}{\partial}
\renewcommand{\Im}{\operatorname{Im}\nolimits}
\renewcommand{\Re}{\operatorname{Re}\nolimits}
\newcommand{\Tr}{\operatorname{Tr}}

\newcommand{\CC}{\mathbf{C}}

\newcommand{\RR}{\mathbf{R}}
\newcommand{\ZZ}{\mathbf{Z}}
\newcommand{\bst}{\boldsymbol{t}}
\newcommand{\bsT}{\boldsymbol{T}}
\newcommand{\bszero}{\boldsymbol{0}}
\newcommand{\calB}{\mathcal{B}}
\newcommand{\calD}{\mathcal{D}}
\newcommand{\calE}{\mathcal{E}}

\newcommand{\calL}{\mathcal{L}}
\newcommand{\calM}{\mathcal{M}}
\newcommand{\calN}{\mathcal{N}}
\newcommand{\calO}{\mathcal{O}}
\newcommand{\calP}{\mathcal{P}}
\newcommand{\fkL}{\mathfrak{L}}
\newcommand{\fkM}{\mathfrak{M}}
\newcommand{\dprime}{\prime\prime}

\begin{document}

\title{Thermodynamic limit of random partitions 
and dispersionless Toda hierarchy}
\author{Kanehisa Takasaki\thanks{takasaki@math.h.kyoto-u.ac.jp}\\
{\normalsize Graduate School of Human and Environmental Studies, 
Kyoto University}\\
{\normalsize Yoshida, Sakyo, Kyoto 606-8501, Japan}
\\[6pt]
Toshio Nakatsu\thanks{nakatsu@mpg.setsunan.ac.jp}\\
{\normalsize Institute for Fundamental Sciences, Setsunan University}\\
{\normalsize Ikeda-Nakamachi, Neyagawa, Osaka 572-8508, Japan}}

\date{}
\maketitle

\begin{abstract}
We study the thermodynamic limit of random partition models 
for the instanton sum of 4D and 5D supersymmetric $U(1)$ 
gauge theories deformed by some physical observables.  
The physical observables correspond to external potentials 
in the statistical model.  The partition function is reformulated 
in terms of the density function of Maya diagrams.  
The thermodynamic limit is governed by a limit shape 
of Young diagrams associated with dominant terms 
in the partition function.  The limit shape is characterized 
by a variational problem, which is further converted 
to a scalar-valued Riemann-Hilbert problem.  
This Riemann-Hilbert problem is solved with the aid 
of a complex curve, which may be thought of 
as the Seiberg-Witten curve of the deformed $U(1)$ gauge theory.  
This solution of the Riemann-Hilbert problem is identified 
with a special solution of the dispersionless Toda hierarchy 
that satisfies a pair of generalized string equations.  
The generalized string equations for the 5D gauge theory are shown 
to be related to hidden symmetries of the statistical model. 
The prepotential and the Seiberg-Witten differential are also 
considered.  
\end{abstract}

\begin{flushleft}
\end{flushleft}

\newpage 

\section{Introduction}

This is a sequel to our previous paper \cite{NNT08} 
on the correlation functions of loop operators 
of the 5D supersymmetric $U(1)$ gauge theory 
in the $\Omega$ background on $\RR^4\times S^1$ 
\cite{Losev-etal03}.   As shown therein, 
a generating function of those correlation functions 
coincides with the partition function of 
the so called ``melting crystal model'' deformed 
by a set of external potentials.  
The melting crystal model is a statistical model 
of random plane partitions (3D Young diagrams).  
Actually, by the technique of diagonal slicing \cite{OR03}, 
the partition function can be converted to 
a sum over ordinary partitions (Young diagrams).  
The latter may be thought of as a $q$-analogue 
of the instanton sum (Nekrasov function) 
of 4D $\calN=2$ supersymmetric gauge theories \cite{Nekrasov02}.  
The undeformed melting crystal model was first proposed 
as a statistical model of the amplitude of 
A-model topological strings \cite{ORV03}.  
A slightly modified model was studied as the instanton sum 
of 5D supersymmetric gauge theories \cite{MNTT04a,MNTT04b}.  
The fully deformed model was introduced 
from the point of view of integrable structures \cite{NT07} 
(see also the review \cite{NT08}). It was, however, 
not clear what physical meaning the external potentials 
have in the context of the 5D gauge theory.  
Our previous paper \cite{NNT08} presented an answer 
to this question.  

Another issue addressed therein is the thermodynamic limit 
of the melting crystal model, from which we attempted 
to derive a 5D analogue of the Seiberg-Witten theory \cite{SW94}. 
In the case of 4D $\calN=2$ supersymmetric gauge theories, 
such a ``microscopic'' derivation was first achieved 
by Nekrasov and Okounkov \cite{NO03}, and later extended 
to deformed models by Nekrasov and Marshakov \cite{MN06,Marshakov07}.  
A clue to this problem is the notion of ``limit shape''.  
In the thermodynamic limit, the partition function 
is dominated by Young diagrams of macroscopic sizes and 
a particular (rescaled) shape.  This limit shape 
can be determined by a variational problem. 
The variational problem, in turn, can be converted 
to a Riemann-Hilbert problem for a complex analytic function. 
Solving the Riemann-Hilbert problem, 
Nekrasov and Marshakov \cite{MN06,Marshakov07} could 
derive the Seiberg-Witten theory in the presence 
of external potentials (which they identified with 
local observables $\Tr\phi^{k+1}/(k+1)$, $k = 1,2,\cdots$, 
in the $\calN=2$ chiral multiplet).  
Inspired by their work, we studied the thermodynamic limit 
of the deformed melting crystal model, 
using a slightly different method that was developed 
for the undeformed model \cite{MNTT04b,Maeda-Nakatsu06}.  
Unfortunately, our calculations for the deformed model 
were wrong; we considered an inappropriate 
Riemann-Hilbert problem, which led to wrong results. 

Our primary problem in this paper is, therefore, 
to formulate a correct Riemann-Hilbert problem 
for the deformed melting crystal model.  
We address this problem in a wider context. 
Namely, we treat both the 4D and 5D theories 
by the same method, thereby hoping to compare 
these two cases on a common ground.  Moreover, 
this enables us, simultaneously, to compare our method 
with the method of Nekrasov and Marshakov. 
As it turns out, one can formulate (and solve) 
an appropriate Riemann-Hilbert problem for a function $W(z)$ 
that amounts to the so called ``loop amplitude'' or 
``resolvent function'' in the theory of random matrices.
The solution is accompanied with a complex curve 
(a double covering of the $z$-plane), which may be 
thought of as the Seiberg-Witten curve of 
the $U(1)$ gauge theory.  In the case of the deformed 4D theory, 
this result fully agrees with the result of 
Nekrasov and Marshakov. 

We further present an interpretation of these calculations 
in the language of the dispersionless Toda hierarchy 
\cite{TT91,TT93}.  The dispersionless Toda hierarchy 
is a long-wave limit of the Toda hierarchy \cite{UT84} 
(see the review \cite{TT95} for more details). 
It is known \cite{NO03,Marshakov07,NT07} 
that both the 4D and 5D Nekrasov functions are related 
to the 1D Toda hierarchy (higher time evolutions 
of the Toda lattice), hence to the Toda hierarchy 
(higher time evolutions of the 2D Toda fields) as well.  
On the other hand, as already observed 
by Marshakov and Nekrasov (loc. cit.), 
the thermodynamic limit corresponds to the long-wave limit 
of the Toda lattice, and the solution of the aforementioned 
Riemann-Hilbert problem in the 4D case gives a solution 
of the dispersionless 1D Toda hierarchy.  
We consider this issue from a different point of view, 
namely, the notion of ``generalized string equations'' 
in the Toda and dispersionless Toda hierarchies. 
This notion stems from string theories 
\cite{EK94,Takasaki94,NTT95,KO95,Takasaki96}, 
and found applications in interface dynamics 
\cite{WZ00,MWWZ00,Zabrodin01}, which eventually led 
to a detailed description of a quite general class 
of solutions of the dispersionless Toda hierarchy \cite{Teo09}.  
Employing this notion and techniques developed 
in earlier studies, we show that the solutions 
of the Riemann-Hilbert problem for both 4D and 5D cases 
can be (almost uniquely) characterized 
by a pair of generalized string equations.  
``Almost'' means that the generalized string equations 
for the 5D case are suffered from a kind of ambiguity. 
The meaning of this ambiguity is left for a future study.  

Let us mention that the notion of generalized string equations 
is also useful for a statistical model of 
the double Hurwitz numbers \cite{Takasaki10}. 
Remarkably, this statistical model and the melting crystal model 
have a common algebraic structure in their hidden symmetries. 
Following this analogy, one an derive 
the generalized string equations of the 5D theory 
from those hidden symmetries of the partition functions.  
We discuss this issue in the end of the paper.  

This paper is organized as follows.  
Sections 2, 3 and 4 show the formulation 
of the random partition models, 
the prescription of thermodynamic limit and 
the derivation of Riemann-Hilbert problems.  
In Section 2, the deformed random partition models 
are formulated.  A 2D free fermion system is used 
to formulate the main statistical weights and 
the external potentials in a compact form.  
In Section 3, the partition functions are reformulated 
in terms of the density function of Maya diagrams.  
The energy functionals for the density function are specified.  
In Section 4, the thermodynamic limit is formulated 
and the asymptotic form of the energy functionals is determined.  
The variational equation for the limit shape is derived 
and converted to a Riemann-Hilbert problem.  
Sections 5, 6, 7 and 8 present the main results 
on solutions of the Riemann-Hilbert problem and 
their relation to the dispersionless Toda hierarchy.  
The solution for the 4D theory is constructed 
in Section 5, and examined in Section 6 
from the point of view of generalized string equations.  
The solution for the 5D theory is considered 
in the same way in Sections 7 and 8.  
Section 8 also presents a derivation of 
the generalized string equations of from hidden symmetries 
of the melting crystal model.  
Section 9 is devoted to concluding remarks 
on the prepotential and the Seiberg-Witten differential.

\section{Random partition models of $U(1)$ gauge theories}

\subsection{Fermions}

We use the same formulation of fermions as 
our previous papers \cite{NT07,NT08,NNT08}. 
Let $\psi_n,\psi^*_n$, $n \in \ZZ$, denote 
the Fourier modes of 2D complex fermion fields 
\beqnn
  \psi(z) = \sum_{n\in\ZZ}\psi_nz^{-n-1},\quad
  \psi^*(z) = \sum_{n\in\ZZ}\psi^*_nz^{-n}. 
\eeqnn
They satisfy anti-commutation relations
\beqnn
  \psi_m\psi^*_n + \psi^*_m\psi_n = \delta_{m+n,0},\quad 
  \psi_m\psi_n + \psi_m\psi_n 
  = \psi^*_m\psi^*_n + \psi^*_n\psi^*_m = 0. 
\eeqnn
The charge-$s$ sector of the Fock space 
has the ground states 
\beqnn
  \langle s| = \langle-\infty|\cdots\psi^*_{s-1}\psi^*_s, \quad 
  |s\rangle = \psi_{-s}\psi_{-s+1}\cdots|-\infty\rangle
\eeqnn
and excited states 
\beqnn
  \langle\mu,s| = \langle -\infty|\cdots\psi^*_{\mu_2+s-1}\psi^*_{\mu_1+s},\quad
  |\mu,s\rangle = \psi_{-\mu_1-s}\psi_{-\mu_2-s+1}\cdots|-\infty\rangle 
\eeqnn
labelled by the set $\calP$ of 
all partitions $\mu = (\mu_i)_{i=1}^\infty$, 
$\mu_1 \ge \mu_2 \ge \cdots \ge 0$, of arbitrary lengths.  

Let us introduce the fermion bilinears 
\beqnn
\begin{gathered}
  J_k = \sum_{n\in\ZZ}{:}\psi_{k-n}\psi^*_n{:},\\
  L_0 = \sum_{n\in\ZZ}n{:}\psi_{-n}\psi^*_n{:},\quad
  W_0 = \sum_{n\in\ZZ}n^2{:}\psi_{-n}\psi^*_n{:}. 
\end{gathered}
\eeqnn
$J_k$'s span a $U(1)$ current algebra, 
and $L_0$ and $W_0$ are zero-modes of 
Virasoro and $W^{(3)}$ algebras. 
$J_0$, $L_0$ and $W_0$ are diagonal with respect to 
the orthonormal basis $\langle\mu,s|$, $|\mu,s\rangle$. 
The diagonal elements can be calculated as 
\beq
\begin{gathered}
  \langle\mu,s|J_0|\mu,s\rangle = s,\\
  \langle\mu,s|L_0|\mu,s\rangle = |\mu| + \frac{s(s+1)}{2},\\
  \langle\mu,s|W_0|\mu,s\rangle
    = \kappa(\mu) + (2p+1)|\mu| + \frac{s(s+1)(2s+1)}{6}, 
\end{gathered}
\eeq
where
\beqnn
  |\mu| = \sum_{i\ge 1}\mu_i, \quad 
  \kappa(\mu) = \sum_{i\ge 1}\mu_i(\mu_i - 2i + 1). 
\eeqnn

We use this fermionic language to formulate 
the random partition models of instanton sums 
for the 4D and 5D $U(1)$ gauge theories.  
The models are deformed by an infinite number 
of external potentials $\Phi_k(\mu,s)$, $k = 1,2,\cdots$.  
The partition function of each model there becomes 
a function $Z_s(\bst)$ of the coupling constants  
$\bst = (t_1,t_2,\cdots)$.  The coupling constants 
play the role of time variables 
in an underlying integrable hierarchy.

\subsection{Partition function for 4D theory}

We start from the the fermionic formula 
\beq
  Z^{\mathrm{4D}}_s(\bst) 
  = \langle s|e^{\hbar^{-1}J_1}\Lambda_0^{2L_0}
    e^{H(\bst)}e^{\hbar^{-1}J_{-1}}|s\rangle,  
\label{4D-Z-fermion}
\eeq
where $\Lambda_0$ and $\hbar$ are positive constants, 
and $H(\bst)$ is a linear combination 
\beqnn
  H(\bst) 
  = \sum_{k=1}^\infty t_kH_k 
\eeqnn
of the zero-modes 
\beqnn
  H_k 
  = \frac{\hbar^{k+1}}{k+1}\sum_{n\in\ZZ}
    (n^{k+1} - (n-1)^{k+1}){:}\psi_{-n}\psi^*_n{:}
\eeqnn
of (a slightly modified version of) 
the ordinary $W_\infty$ algebra.   

Since the action of $e^{J_1}$, $\Lambda_0^{2L_0}e^{H(\bst)}$ 
and $e^{J_{-1}}$ preserves the charge-$s$ sector, 
one can insert the partition of unity 
\beqnn
  1 = \sum_{\mu\in\calP}|\mu,s\rangle\langle\mu,s| 
  \quad \text{(in charge-$s$ sector)}
\eeqnn
between these operators, where $\calP$ denotes 
the set of all partitions of arbitrary length. 
This leads to an expansion of $Z^{\mathrm{4D}}_s(\bst)$ 
into a sum over partitions.  Moreover, 
since $\Lambda^{2L_0}e^{H(\bst)}$ is diagonal, 
this becomes a single (rather than double) sum:
\beqnn
  Z^{\mathrm{4D}}_s(\bst) 
  = \sum_{\mu\in\calP}
    \langle s|e^{\hbar^{-1}J_1}|\mu,s\rangle
    \langle\mu,s|\Lambda_0^{2L_0}e^{H(\bst)}|\mu,s\rangle
    \langle\mu,s|e^{\hbar^{-1}J_{-1}}|s\rangle. 
\eeqnn

Let us specify the building blocks of this sum.  
Firstly, $e^{J_{\pm 1}}$ are known to act 
on the ground states as 
\beq
  \langle s|e^{\hbar^{-1}J_1} 
  = \sum_{\mu\in\calP}\langle\mu,s|\frac{\dim\mu}{\hbar^{|\mu|}|\mu|!},\quad
  e^{\hbar^{-1}J_{-1}}|s\rangle
  = \sum_{\mu\in\calP}\frac{\dim\mu}{\hbar^{|\mu|}|\mu|!}|\mu,s\rangle, 
\eeq
where $\dim\mu$ denotes the degree $\dim S^\mu$ 
of the irreducible representation $S^\mu$ of 
the symmetric group $S_r$, $r = |\mu|$ \cite{Sagan-book}. 
$\dim\mu$ has the hook length formula 
\beq
   \dim\mu = |\mu|!\prod_{\square\in\mu}h(\square)^{-1}, 
\label{dim-mu-formula}
\eeq
where $h(\square)$ denotes the hook length of the cell $\square$ 
in the Young diagram of $\mu$.  Secondly, the diagonal elements 
of $\Lambda^{2L_0}e^{H(\bst)}$ can be factorized as 
\beq
  \langle\mu,s|\Lambda_0^{2L_0}e^{H(\bst)}|\mu,s\rangle 
  = \Lambda_0^{2|\mu|+s(s+1)}e^{\Phi(\bst,\mu,s)}, 
\eeq
where 
\beqnn
  \Phi(\bst,\mu,s) = \sum_{k=1}^\infty t_k\Phi_k(\mu,s),\quad 
  \Phi_k(\mu,s) = \langle\mu,s|H_k|\mu,s\rangle. 
\eeqnn
$\Phi_k(\mu,s)$ has the formal (apparently divergent) expression 
\beq
  \Phi_k(\mu,s)
  = \frac{\hbar^{k+1}}{k+1}\sum_{i=1}^\infty
     \left((s+\mu_i-i+1)^{k+1} - (s+\mu_i-i)^{k+1}\right), 
\eeq
which can be reorganized to a finite sum as 
\begin{multline}
  \Phi_k(\mu,s) 
  = \frac{\hbar^{k+1}}{k+1}\sum_{i=1}^\infty
     \left((s+\mu_i-i+1)^{k+1} - (s-i+1)^{k+1}\right)\\
  \mbox{}
    - \frac{\hbar^{k+1}}{k+1}\sum_{i=1}^\infty
      \left((s+\mu_i-i)^{k+1} - (s-i)^{k+1}\right) 
    + \frac{(\hbar s)^{k+1}}{k+1}. 
\end{multline}

Thus $Z^{\mathrm{4D}}_s(\bst)$ can be converted to the sum 
\beq
  Z^{\mathrm{4D}}_s(\bst) 
  = \sum_{\mu\in\calP}
      \left(\frac{\dim\mu}{\hbar^{|\mu|}|\mu|!}\right)^2
      \Lambda_0^{2|\mu|+s(s+1)}e^{\Phi(\bst,\mu,s)}
\label{4D-Z-partition}
\eeq
over all partitions.  This is exactly the Nekrasov function 
of the 4D $\calN=2$ supersymmetric $U(1)$ gauge theory 
deformed by external potentials \cite{MN06,Marshakov07}.

\subsection{Partition function for 5D theory}

We define a 5D analogue of (\ref{4D-Z-fermion}) as 
\beq
  Z^{\mathrm{5D}}_s(\bst) 
  = \langle s|G_{+}Q^{L_0}e^{H(\bst)}G_{-}|s\rangle, 
\label{5D-Z-fermion}
\eeq
where $q$ and $Q$ are constant with $0 < q < 1$ and $0 < Q < 1$, 
$H(\bst)$ is a linear combination 
\beqnn
  H(\bst) = \sum_{k=1}^\infty t_kH_k 
\eeqnn
of the zero-modes 
\beqnn
  H_k = \sum_{n\in\ZZ}q^{kn}{:}\psi_{-n}\psi^*_n{:} 
\eeqnn
of the quantum-torus algebra \cite{NT08}, 
and $G_{\pm}$ are the transfer operators 
\beqnn
  G_{\pm} = \exp\left(\sum_{k=1}^\infty
               \frac{q^{k/2}}{k(1-q^k)}J_{\pm k}\right)
\eeqnn
of Okounkov and Reshetikhin \cite{OR03}.  

Inserting the partition of unity between operators, 
one can expand $Z^{\mathrm{5D}}_s(\bst)$ 
into a sum over all partitions.  The building blocks 
of this expansion can be calculated as follows.  
According to Okounkov and Reshetikhin \cite{OR03}, 
$G_{\pm}$ act on the ground states $\langle s|$, $|s\rangle$ as 
\beq
  \langle s|G_{+} = \sum_{\mu\in\calP}\langle\mu,s|s_\mu(q^\rho),\quad 
  G_{-}|s\rangle = \sum_{\mu\in\calP}s_\mu(q^\rho)|\mu,s\rangle, 
\eeq
where $s_\mu(q^\rho)$ is the special value 
of the Schur function $s_\mu(x_1,x_2,\cdots)$ 
of infinite variables \cite{Macdonald-book} at 
\beqnn
  \rho = (q^{1/2},q^{3/2},\cdots,q^{n-1/2},\cdots).  
\eeqnn
This special value has the hook length formula 
\beq
  s_\mu(q^\rho) 
   = q^{-\kappa(\mu)/4}
     \prod_{\square\in\mu}(q^{-h(\square)/2} - q^{h(\square)/2})^{-1}, 
\label{s_mu(q^rho)-formula}
\eeq
which may be thought of as a $q$-deformation 
of the hook length formula (\ref{dim-mu-formula}) of $\dim\mu$.  
The operator $Q^{L_0}e^{H(\bst)}$ in the middle is diagonal, 
and the diagonal elements can be factorized as 
\beq
  \langle\mu,s|Q^{L_0}e^{H(\bst)}|\mu,s\rangle 
  = Q^{|\mu|+s(s+1)/2}e^{\Phi(\bst,\mu,s)}, 
\eeq
where 
\beqnn
  \Phi(\bst,\mu,s) = \sum_{k=1}^\infty t_k\Phi_k(\mu,s),\quad 
  \Phi_k(\mu,s) = \langle\mu,s|H_k|\mu,s\rangle. 
\eeqnn
$\Phi_k(\mu,s)$ has the formal expression (apparently valid 
for $|q| > 1$) 
\beq
  \Phi_k(\mu,s) 
  = \sum_{i=1}^\infty q^{k(s+\mu_i-i+1)} 
    - \sum_{i=1}^\infty q^{k(-i+1)}, 
\eeq
which can be reorganized to a finite sum as 
\beq
  \Phi_k(\mu,s) 
  = \sum_{i=1}^\infty(q^{k(s+\mu_i-i+1)} - q^{k(s-i+1)}) 
    + \frac{q^k(1-q^{ks})}{1-q^k}. 
\eeq

Thus $Z^{\mathrm{5D}}_s(\bst)$ can be expanded to 
a sum over all partitions as 
\beq
  Z^{\mathrm{5D}}_s(\bst) 
  = \sum_{\mu\in\calP}
     s_\mu(q^\rho)^2Q^{|\mu|+s(s+1)/2}e^{\Phi(\bst,\mu,s)}. 
\label{5D-Z-partition}
\eeq
This is a 5D analogue of the deformed 4D $U(1)$ Nekrasov function 
(\ref{4D-Z-partition}). Although one can further insert 
a Chern-Simons term \cite{MNTT04a}, 
we consider this simplest model in this paper.

\section{Energy functional for density function}

\subsection{Density function of Maya diagram}

We now recall the notion of density function 
of Maya diagrams \cite{MNTT04b,Maeda-Nakatsu06} 
that play a central role in our approach 
to the thermodynamic limit.  

The Maya diagram of $\mu$ is a configuration of particles 
located at the points $x = \mu_i - i$, $i = 1,2,\cdots$, on a line. 
Let $\rho_\mu(x)$ denote the density function 
\beqnn
  \rho_\mu(x) = \sum_{i=1}^\infty \delta(x - \mu_i + i)
\eeqnn
of this particle configuration.  The backward difference
\beqnn
  \Delta\rho_{\mu}(x) = \rho_\mu(x) - \rho_\mu(x-1)
\eeqnn
becomes, up to a constant factor, the second derivative 
of the so called ``profile function'' $f_\mu(x)$ \cite{NO03} 
of the ($45$ degrees rotated) Young diagram: 
\beq
  \Delta\rho_\mu(x) = - \frac{1}{2}f_\mu''(x).  
\eeq
The first few moments of $\Delta\rho_\mu(x)$ 
can be explicitly calculated as 
\beq
\begin{gathered}
  \int_{-\infty}^\infty dx\Delta\rho_\mu(x) = -1,\\ 
  \int_{-\infty}^\infty dx x\Delta\rho_\mu(x) = 0,\\
  \int_{-\infty}^\infty dx x^2\Delta\rho_\mu(s) = -2|\mu|,\\
  \int_{-\infty}^\infty dx x^3\Delta\rho_\mu(s) = -3\kappa(\mu).
\end{gathered}
\label{Drho-moment}
\eeq

The density function $\rho_{\mu,s}(x)$ 
for the charged partition $(\mu,s)$ 
is a parallel transform of $\rho_\mu(x)$: 
\beqnn
  \rho_{\mu,s}(x) 
  = \rho_\mu(x-s) 
  = \sum_{i=1}^\infty \delta(x - s - \mu_i + i). 
\eeqnn
The first few moments read
\beq
\begin{gathered}
  \int_{-\infty}^\infty dx\Delta\rho_{\mu,s}(x) = -1, \\
  \int_{-\infty}^\infty dx x\Delta\rho_{\mu,s}(x) = -s,\\
  \int_{-\infty}^\infty dx x^2\Delta\rho_{\mu,s}(s) 
     = -2|\mu| - s^2,\\
  \int_{-\infty}^\infty dx x^3\Delta\rho_{\mu,s}(s) 
     = -3\kappa(\mu) - 6s|\mu| - s^3. 
\end{gathered}
\label{Drho_s-moment}
\eeq

\subsection{External potentials in terms of density function}

Let $\calO^{\mathrm{4D}}_k(\mu,s)$ denote the $k$-th moment 
of $-\Delta\rho_{\mu,s}(x)$: 
\beqnn
  \calO^{\mathrm{4D}}_k(\mu,s) 
  = - \int_{-\infty}^\infty dx x^k\Delta\rho_{\mu,s}(x).
\eeqnn
These moments have a formal expression 
\beq
  \calO^{\mathrm{4D}}_k(\mu,s) 
  = \sum_{i=1}^\infty (s+\mu_i-i+1)^k 
   - \sum_{i=1}^\infty (s+\mu_i-i)^k, 
\eeq
which can be converted to a finite sum as 
\begin{multline}
  \calO^{\mathrm{4D}}_k(\mu,s) 
  = \sum_{i=1}^\infty\left((s+\mu_i-i+1)^k - (s-i+1)^k \right)\\
  \mbox{} 
   - \sum_{i=1}^\infty\left((s+\mu_i-i)^k - (s-i)^k\right) 
   + s^k. 
\end{multline}
The external potentials 
\beqnn
  \Phi^{\mathrm{4D}}_k(\mu,s)
  = \frac{\hbar^{k+1}}{k+1}\sum_{i=1}^\infty
     \left((s+\mu_i-i+1)^{k+1} - (s+\mu_i-i)^{k+1}\right) 
\eeqnn 
of the 4D gauge theory can be thereby expressed as 
\beq
  \Phi^{\mathrm{4D}}_k(\mu,s) 
  = \frac{\hbar^{k+1}\calO^{\mathrm{4D}}_{k+1}(\mu,s)}{k+1}
  = - \int_{-\infty}^\infty dx
      \frac{(\hbar x)^{k+1}}{k+1}\Delta\rho_{\mu,s}(x).
\label{4D-Phi-rho}
\eeq

One can find a similar expression of the external potentials 
for the 5D gauge theory.  Consider the $q$-analogue 
\beqnn
  \calO^{\mathrm{5D}}_k(\mu,s) 
  = - \int_{-\infty}^\infty dx q^{kx}\Delta\rho_{\mu,s}(x) 
\eeqnn
of $\calO^{\mathrm{4D}}_k(\mu,s)$.  The formal expression 
\beq
  \calO^{\mathrm{5D}}_k(\mu,s) 
  = \sum_{i=1}^\infty q^{k(s+\mu_i-i+1)} 
     - \sum_{i=1}^\infty q^{k(s+\mu_i-i)} 
\eeq
can be converted to a finite sum as 
\begin{multline}
  \calO^{\mathrm{5D}}_k(\mu,s)
  = \sum_{i=1}^\infty\left(q^{k(s+\mu_i-i+1)} - q^{k(s-i+1)}\right)\\
  \mbox{} 
   - \sum_{i=1}^\infty\left(q^{k(s+\mu_i-i)} - q^{k(s-i)}\right)
   + q^{ks}. 
\end{multline}
The external potentials 
\beqnn
    \Phi^{\mathrm{5D}}_k(\mu,s) 
  = \sum_{i=1}^\infty q^{k(s+\mu_i-i+1)} 
    - \sum_{i=1}^\infty q^{k(-i+1)} 
\eeqnn
of the 5D gauge theory are related to these $q$-moments as 
\beq
\begin{aligned}
  \Phi^{\mathrm{5D}}_k(\mu,s) 
  &= - \frac{q^k}{1-q^k}\calO^{\mathrm{5D}}_k(\mu,s) + \frac{q^k}{1-q^k}\\
  &= \frac{q^k}{1-q^k}\int_{-\infty}^\infty 
     dx q^{kx}\Delta\rho_{\mu,s}(x)
    + \frac{q^k}{1-q^k}. 
\end{aligned}
\label{5D-Phi-rho}
\eeq

\subsection{Hook product in terms of density function}

By the hook length formulae (\ref{dim-mu-formula}) and 
(\ref{s_mu(q^rho)-formula}), the main part of the Boltzmann weights 
in (\ref{4D-Z-partition}) and (\ref{5D-Z-partition}) 
can be rewritten as 
\begin{gather}
  \left(\frac{\dim\mu}{\hbar^{|\mu|}|\mu|!}\right)^2\Lambda_0^{2|\mu|}
    = \prod_{\square\in\mu}\left(\frac{\Lambda_0}{\hbar}\right)^2h(\square)^{-2},
  \label{4D-weight}\\
  s_\mu(q^\rho)^2Q^{|\mu|}
    = q^{-\kappa(\mu)/2}\prod_{\square\in\mu}Qq^{h(\square)}(1 - q^{h(\square)})^{-2}. 
  \label{5D-weight}
\end{gather}
Let us recall the general formula \cite{NNT08} 
\beq
  \sum_{\square\in\mu}f(h(\square))
  = \frac{1}{2}\iint_{-\infty}^\infty 
    dxdy g(|x-y|)\Delta\rho_{\mu}(x)\Delta\rho_\mu(y), 
\label{cell-sum-formula}
\eeq
where $f(x)$ is an arbitrary function and $g(x)$ 
(referred to as the kernel function) is a function 
that satisfies the conditions 
\beq
  g(x+1) + g(x-1) - 2g(x) = f(x), \quad 
  g(0) = 0. 
\eeq
By this formula, the logarithm of (\ref{4D-weight}) 
and (\ref{5D-weight}) can be converted to quadratic functionals 
of $\Delta\rho_{\mu,s}$ as follows. 
\begin{itemize}
\item 4D theory \cite{NO03}:  One first obtains the expression 
\beqnn
  - \log \left(\frac{\dim\mu}{\hbar^{|\mu|}|\mu|!}\right)^2\Lambda_0^{2|\mu|}
  = \iint_{-\infty}^\infty dxdy 
    g_{\mathrm{4D}}(|x-y|)\Delta\rho_\mu(x)\Delta\rho_\mu(y), 
\eeqnn
where 
\beqnn
  g_{\mathrm{4D}}(x) 
  = -\frac{x(x-1)}{2} \log \frac{\Lambda_0}{\hbar} + \log G_2(x+1). 
\eeqnn
$G_2(x)$ is Barnes' G-function, 
the second member of the multiple Gamma functions $G_n(x)$,\,$n=0,1,\cdots$.
Note that $g_{\mathrm{4D}}(x)$ satisfies the equation 
\beq
  g(x+1) + g(x-1) - 2g(x) = \log\frac{\hbar x}{\Lambda_0}. 
\eeq
Substituting $x \to x - s$, one can rewrite $\Delta\rho_\mu$ 
to $\Delta\rho_{\mu,s}$ as 
\beq
  - \log \left(\frac{\dim\mu}{\hbar^{|\mu|}|\mu|!}\right)^2\Lambda_0^{2|\mu|}
  = \iint_{-\infty}^\infty dxdy 
    g_{\mathrm{4D}}(|x-y|)\Delta\rho_{\mu,s}(x)\Delta\rho_{\mu,s}(y), 
\label{4D-log-weight}
\eeq
\item 5D theory \cite{NNT08}:  Also using the last formula 
of (\ref{Drho-moment}) to the term $(\kappa(\mu)/2)\log q$, 
one first obtains the expression 
\begin{multline*}
  - \log s_\mu(q^\rho)^2Q^{|\mu|}
  = \iint_{-\infty}^\infty dxdy g_{\mathrm{5D}}(|x-y|)
       \Delta\rho_\mu(x)\Delta\rho_\mu(y)\\
  \mbox{} - \frac{\log q}{6}\int_{-\infty}^\infty dx x^3\Delta\rho_\mu(x),
\end{multline*}
where
\begin{multline*}
  g_{\mathrm{5D}}(x) 
  = - \frac{x(x-1)}{4}\log Q - \frac{x(x^2-1)}{12}\log q \\
    \mbox{} + \frac{x(x-1)}{2}\log(1-q) + \log G_2(x+1;q). 
\end{multline*}
$G_2(x;q)$ is the second member of the multiple $q$-Gamma functions 
$G_n(x;q)$, $n = 0,1,\cdots$ \cite{Nishizawa95}.  
Note that $g_{\mathrm{5D}}(x)$ is chosen to satisfy the equation 
\begin{multline}
  g(x+1) + g(x-1) - 2g(x) 
  =  - \frac{\log Q}{2} - \frac{x\log q}{2} 
     + \log(1 - q^x). 
\end{multline}
Substituting $x \to x - s$, one can rewrite $\Delta\rho_\mu$ 
to $\Delta\rho_{\mu,s}$ as 
\beq
  \begin{aligned}
  - \log s_\mu(q^\rho)^2Q^{|\mu|}
  &= \iint_{-\infty}^\infty dxdy g_{\mathrm{5D}}(|x-y|)
       \Delta\rho_{\mu,s}(x)\Delta\rho_{\mu,s}(y)\\
  &\quad - \frac{\log q}{6}\int_{-\infty}^\infty dx (x-s)^3\Delta\rho_{\mu,s}(x),
  \end{aligned}
\label{5D-log-weight}
\eeq
\end{itemize}

\subsection{Energy functional for density function}

Thus the main part of the Boltzmann weight becomes 
the exponential of a quadratic functional of $\Delta\rho_{\mu,s}$.  
The external potentials are already shown 
to have such a functional expression by 
(\ref{4D-Phi-rho}) and (\ref{5D-Phi-rho}).  
Thus, up to a simple factor independent of $\Delta\rho_{\mu,s}$, 
the Boltzmann weight can be cast into 
the standard form $e^{-\calE[\rho_{\mu,s}]}$ 
with the following energy functionals $\calE[\rho]$: 
\begin{itemize}
\item 4D theory:  
\begin{multline}
  \calE^{\mathrm{4D}}_{\bst}[\rho] 
  = \iint_{-\infty}^\infty dxdy
    g_{\mathrm{4D}}(|x-y|) \Delta\rho(x)\Delta\rho(y) \\
  \mbox{} 
    + \int_{-\infty}^\infty dx
      \left(\sum_{k=1}^\infty\frac{t_k(\hbar x)^{k+1}}{k+1}\right)
      \Delta\rho(x).
\label{4D-E-functional}
\end{multline}
\item 5D theory: 
\begin{multline}
  \calE^{\mathrm{5D}}_{s,\bst}[\rho]
  = \iint_{-\infty}^\infty dxdy 
      g_{\mathrm{5D}}(|x-y|)\Delta\rho(x)\Delta\rho(y)\\
  \mbox{} - \frac{\log q}{6}
      \int_{-\infty}^\infty dx(x-s)^3\Delta\rho(x) 
    - \int_{-\infty}^\infty dx 
      \left(\sum_{k=1}^\infty\frac{t_kq^k}{1-q^k}q^{kx}\right)
      \Delta\rho(x). 
\label{5D-E-functional}
\end{multline}
\end{itemize}
We now have the following expression of the partition functions 
as statistical sums over the set $\calD_s$ of all density functions 
of the form $\rho = \rho_{\mu,s}$, $\mu \in \calP$: 
\begin{gather}
  Z^{\mathrm{4D}}_s(\bst) 
  = \Lambda_0^{s(s+1)}\sum_{\rho\in\calD_s}
    e^{-\calE^{\mathrm{4D}}_{\bst}[\rho]},
  \label{4D-Z-rho}\\
  Z^{\mathrm{5D}}_s(\bst) 
  = \exp\left(\sum_{k=1}^\infty\frac{t_kq^k}{1-q^k}\right)
    Q^{s(s+1)/2}
    \sum_{\rho\in\calD_s}e^{-\calE^{\mathrm{5D}}_{s,\bst}[\rho]}.
  \label{5D-Z-rho}
\end{gather}

\section{Thermodynamic limit and Riemann-Hilbert problem}

\subsection{Thermodynamic limit for 4D theory}

To achieve the thermodynamic limit for the 4D theory, 
we rescale $s$ and $t_k$'s as 
\beq
  s \to \hbar^{-1}s, \quad 
  t_k \to \hbar^{-2}t_k,
\label{4D-rescaling}
\eeq
and let $\hbar \to 0$.  The partition function 
is then dominated by partitions $\mu$ 
with $|\mu| = O(\hbar^{-2})$. 
The rescaled profile functions 
\beqnn
  \hbar f_{\mu,\hbar^{-1}s}(\hbar^{-1}u) 
  = \hbar f_\mu(\hbar^{-1}(u-s)) 
\eeqnn
of these dominant partitions then tend to 
a continuous ``limit shape'' $f^{(0)}_*(u)$. 

To formulate this limit shape in the language 
of the density functions of Maya diagrams, 
we assume that the dominant contribution 
to the partition function is due to 
density functions of the form 
\beq
  \rho(\hbar^{-1}u) = \rho^{(0)}(u) + O(\hbar) 
  \quad \text{as $\hbar \to 0$}, 
\eeq
where $\rho^{(0)}(u)$ is a continuous function, 
monotonously decreasing, and differentiable 
in a {\it weak sense\/}, 
\footnote{This means that 
the derivative $\rho^{(0)\prime}$ can be singular 
at some exceptional points, but that the various integrals 
containing $\rho^{(0)\prime}$ are still meaningful.}
and the derivative $\rho^{(0)\prime}(u)$ satisfies 
the constraints 
\beq
  \int_{-\infty}^\infty du\rho^{(0)\prime}(u) = -1,\quad
  \int_{-\infty}^\infty duu\rho^{(0)\prime}(u) = - s. 
\label{rho0'-constraint}
\eeq
Note that these constraints stem from the moment formula 
(\ref{Drho_s-moment}). 

Under these assumptions, we can find the leading part 
(of order $\hbar^{-2}$) of the energy functional 
$\calE^{\mathrm{4D}}_{s\bst}[\rho]$ explicitly. 
Note that the kernel function $g_{\mathrm{4D}}$ 
in the rescaled coordinate behaves as 
\beq
  g_{\mathrm{4D}}(\hbar^{-1}u) 
  = \hbar^{-2}g_{\mathrm{4D}}^{(0)}(u) + O(\hbar^{-1}), 
\eeq
where
\beqnn
  g_{\mathrm{4D}}^{(0)}(u) 
  = \frac{u^2}{2}\left(\log\frac{u}{\Lambda_0} - \frac{3}{2}\right). 
\eeqnn
Also note that $\Delta\rho(x)$ in the rescaled coordinate 
can be expressed as 
\beq
  \Delta\rho(\hbar^{-1}u) = \hbar\rho^{(0)\prime}(u) + O(\hbar^2). 
\eeq
Consequently, the quadratic part of $\calE^{\mathrm{4D}}_{\bst}[\rho]$ 
turns out to have the following asymptotic form: 
\begin{multline*}
  \iint_{-\infty}^\infty dxdyg_{\mathrm{4D}}(|x-y|)
     \Delta\rho(x)\Delta(y) \\
  = \hbar^{-2}\iint_{-\infty}^\infty dudv 
      g_{\mathrm{4D}}^{(0)}(|u-v|)
      \rho^{(0)\prime}(u)\rho^{(0)\prime}(v) 
    + O(\hbar^{-1}).
\end{multline*}
On the other hand, the linear part, after rescaling 
$t_k \to \hbar^{-2}t_k$, behaves as 
\begin{multline*}
  \int_{-\infty}^\infty dx
  \left(\sum_{k=1}^\infty\frac{\hbar^{-2}t_k(\hbar x)^{k+1}}{k+1}\right) 
  \Delta\rho(x) \\
  = \hbar^{-2}\int_{-\infty}^\infty du 
    \left(\sum_{k=1}^\infty\frac{t_ku^{k+1}}{k+1}\right)
    \rho^{(0)\prime}(u) 
    + O(\hbar^{-1}). 
\end{multline*}
Thus the energy functional itself can be expressed as 
\beq
  \calE^{\mathrm{4D}}_{\hbar^{-2}\bst}[\rho]
  = \hbar^{-2}\calE^{\mathrm{4D}(0)}_{\bst}[\rho^{(0)}] 
    + O(\hbar^{-1}),
\eeq
where 
\begin{multline}
  \calE^{\mathrm{4D}(0)}_{\bst}[\rho^{(0)}] 
  = \iint_{-\infty}^\infty dudvg_{\mathrm{4D}}^{(0)}(|u-v|)
     \rho^{(0)\prime}(u)\rho^{(0)\prime}(v) \\
  \mbox{} 
    + \int_{-\infty}^\infty du
      \left(\sum_{k=1}^\infty\frac{t_ku^{k+1}}{k+1}\right)\rho^{(0)\prime}(u). 
\end{multline}

\subsection{Riemann-Hilbert problem for 4D theory}

The density function 
\beqnn
  \rho^{(0)}_*(u) = \frac{1 - f^{(0)}_*(u)}{2} 
\eeqnn
of the limit shape is obtained as a minimizer 
of $\calE^{\mathrm{4D}(0)}_{\bst}[\rho^{(0)}]$ 
under the constraints (\ref{rho0'-constraint}).  
Following the well known ansatz, we assume that 
$\rho^{(0)}_*(u)$ is constant outside an interval $[u_0,u_1]$. 
This amounts to the ``one-cut'' condition 
in the large-$N$ limit of random matrix models. 
Since the test functions $\rho^{(0)}(u)$ 
in the minimization problem are monotonously decreasing 
continuous functions with the obvious boundary conditions 
\beq
  \lim_{u\to -\infty}\rho^{(0)}(u) = 1, \quad 
  \lim_{u\to +\infty}\rho^{(0)}(u) = 0, 
\eeq
this means that 
\beq
  \rho^{(0)}_*(u) 
  = \begin{cases}
    1  & \text{for $u < u_0$},\\
    0  & \text{for $u > u_1$}.
    \end{cases}
\label{rho0'-one-cut}
\eeq

If we limit the test functions to those with 
the last property (\ref{rho0'-one-cut}), 
the variational equation (with Lagrange multiplier $\nu$) 
\beqnn
  \frac{\delta}{\delta\rho^{(0)}}
  \left(\calE^{\mathrm{4D}(0)}_{\bst}[\rho^{(0)}] 
    + \nu\left(\int_{-\infty}^\infty duu\rho^{(0)\prime}(u) + s\right) 
  \right) = 0 
\eeqnn
reduces to solving the integral equation 
\beq
\frac{d}{du} \left( \int_{u_0}^{u_1}dv g_{\mathrm{4D}}^{(0)}(|u-v|)
   \rho^{(0)\prime}(v) \right) 
  + \frac{V(u)}{2} + \frac{\nu}{2} 
  = 0 \quad (u_0 \le u \le u_1), 
\label{4D-rho0'-eq}
\eeq
where 
\beqnn
  V(u) = \sum_{k=1}^\infty t_ku^k, 
\eeqnn
under the constraints (\ref{rho0'-constraint}) 
and the support condition 
\footnote{Actually, the derivative $\rho^{(0)\prime}_*(u)$ 
of the minimizer turns out to have singularities 
at the endpoints $u_0,u_1$ of the support.  The integrals 
in (\ref{4D-rho0'-eq}) and (\ref{rho0'-constraint}), 
however, have finite values.}
\beq
  \rho^{(0)\prime}(u) = 0 \quad\text{for $u \not\in [u_0,u_1]$}. 
\eeq
Let us stress that the endpoints $u_0,u_1$ are ``dynamical'', 
namely, determined along with the minimizer $\rho^{(0)\prime}_*(u)$.  

Actually, rather than solving (\ref{4D-rho0'-eq}) directly, 
we consider the equation 
\beq
  \mathrm{PP}\int_{u_0}^{u_1}dv 
   g_{\mathrm{4D}}^{(0)\dprime}(|u-v|)\rho^{(0)\prime}(v)
  + \frac{V'(u)}{2}  
  = 0 \quad (u_0 \le u \le u_1) 
\label{4D-rho0'-eq1}
\eeq
obtained by differentiating (\ref{4D-rho0'-eq}). 
Note that the integral on the left hand side 
is now interpreted to be the principal part 
(indicated by the notation ``PP''), 
because the kernel function 
\beqnn
  g_{\mathrm{4D}}^{(0)\dprime}(|u-v|) 
  = \log\frac{|u-v|}{\Lambda_0} 
\eeqnn
has a logarithmic singularity at $v = u$.  

We now introduce the complexified kernel function
\beqnn
  g_{\mathrm{4D}}^{(0)\dprime}(z-v) = \log\frac{z-v}{\Lambda_0}. 
\eeqnn
and construct the complex analytic function 
\beqnn
  W(z) = \int_{u_0}^{u_1}dv g_{\mathrm{4D}}^{(0)\dprime}(z-v)
         \rho^{(0)\prime}_*(v) 
\eeqnn
from the derivative $\rho^{(0)\prime}_*(u)$ of the minimizer. 
$W(z)$ is a holomorphic function on the cut-plane 
$\CC \setminus [u_0,u_1]$ with the boundary values 
\beq
  W(u \pm i0) 
  = \mathrm{PP}\int_{u_0}^{u_1}dv 
        g_{\mathrm{4D}}^{(0)\dprime}(|u-v|)\rho^{(0)\prime}_*(v) 
    \mp \pi i\rho^{(0)}_*(u) 
\eeq
along the real axis.  One can thereby rewrite 
(\ref{4D-rho0'-eq1}) as 
\beq
  W(u + i0) + W(u-i0) = - V'(u) \quad (u_0 \le u \le u_1), 
  \label{4D-RH(1)}
\eeq
and supplement this equation with the equation 
\beq
  W(u+i0) - W(u-i0) 
  = \begin{cases} 
    0       & (u > u_1),\\ 
    -2\pi i & (u < u_0)
    \end{cases} 
    \label{4D-RH(2)}
\eeq
off the support of $\rho^{(0)\prime}_*$.  
Moreover, from the asymptotic form 
\beqnn
  g_{\mathrm{4D}}^{(0)\dprime}(z-v) 
  = \log\frac{z}{\Lambda_0} - \frac{v}{z} + O(z^{-2}) 
  \quad \text{as $z \to \infty$}
\eeqnn
of the kernel function and the constraints 
(\ref{rho0'-constraint}), one can see that 
$W(z)$ should satisfy the boundary condition 
\beq
  W(z) = - \log\frac{z}{\Lambda_0} + \frac{s}{z} + O(z^{-2}) 
  \quad \text{as $z \to \infty$}. 
  \label{4D-RH(3)}
\eeq
Conversely, the constraints (\ref{rho0'-constraint}) 
can be recovered from this boundary condition. 

(\ref{4D-RH(1)}), (\ref{4D-RH(2)}) and (\ref{4D-RH(3)}) 
altogether form a kind of Riemann-Hilbert problem.  
Once a solution of this Riemann-Hilbert problem is found, 
the minimizer $\rho^{(0)}_*$ can be obtained 
from the imaginary part of $W(u\pm 0)$ as 
\beq
  \rho^{(0)}_*(u) = \mp\frac{\Im W(u \pm i0)}{\pi}. 
\label{rho0*-ImW}
\eeq

Let us mention that one can translate 
the variational equation (\ref{4D-rho0'-eq}) itself, 
too, to a Riemann-Hilbert problem.  
This Riemann-Hilbert problem is formulated 
for the primitive function 
\footnote{This amounts to the $G$-function 
in the large-$N$ limit of random matrices.} 
\beqnn
  G(z) = \int^z dzW(z) 
\eeqnn
(with suitable normalization of the integration constant) 
of $W(z)$.  One can solve this Riemann-Hilbert problem 
in much the same way as that of $W(z)$. 
Actually, this is exactly how Marshakov and Nekrasov 
treated the thermodynamic limit \cite{MN06,Marshakov07}.  
Speaking more precisely, they use the function 
\beq
  S(z) = 2G(z) + V(z)
\label{S-function}
\eeq
rather than $G(z)$ itself.  This function $S(z)$ 
coincides with the $S$-function in the theory of 
dispersionless integrable hierarchies \cite{TT95}.

\subsection{Thermodynamic limit for 5D theory}

In the case of the 5D theory, we set the parameters 
$q$ and $Q$ in an $\hbar$-dependent form as 
\beq
  q = e^{-R\hbar}, \quad Q = (R\Lambda_0)^2, 
\eeq
where $R$ and $\Lambda_0$ are positive constants 
with $R\Lambda_0 < 1$, and rescale $s$ and $t_k$'s as 
\beq
  s \to \hbar^{-1}s,\quad 
  t_k \to \hbar^{-1}t_k.
\label{5D-rescaling}
\eeq

Assuming the same properties for the rescaled density function 
$\rho(\hbar^{-1}u)$, one can derive an explicit form 
of the thermodynamic energy functional 
$\calE^{\mathrm{5D}(0)}_{s,\bst}[\rho^{(0)}]$.  
Note that the kernel function $g_{\mathrm{5D}}$ 
in the rescaled coordinate behaves as 
\beq
 g_{\mathrm{5D}}(\hbar^{-1}u) 
  = \hbar^{-2}g_{\mathrm{5D}}^{(0)}(u) + O(\hbar^{-1}).
\eeq
$g_{\mathrm{5D}}^{(0)}(u)$ is a somewhat complicated function; 
for our purpose, it is enough to know that this function 
can be characterized by the following conditions: 
\beq
  g_{\mathrm{5D}}^{(0)\dprime}(u) 
  = \log\frac{e^{Ru/2}-e^{-Ru/2}}{R\Lambda_0},\quad
  g_{\mathrm{5D}}^{(0)\prime}(0) = 0, \quad 
  g_{\mathrm{5D}}^{(0)}(0) = 0. 
\eeq
Hence the quadratic part of $\calE^{\mathrm{5D}}_{s,\bst}[\rho]$ 
has the following asymptotic form: 
\begin{multline*}
    \iint_{-\infty}^\infty dxdyg_{\mathrm{5D}}(|x-y|)
     \Delta\rho(x)\Delta(y) \\
  = \hbar^{-2}\iint_{-\infty}^\infty dudv 
      g_{\mathrm{5D}}^{(0)}(|u-v|)
      \rho^{(0)\prime}(u)\rho^{(0)\prime}(v) 
    + O(\hbar^{-1}).
\end{multline*}
On the other hand, the building blocks 
of the linear part behave as 
\beqnn
  - \frac{\log q}{6}\int_{-\infty}^\infty dx (x - \hbar^{-1}s)^3\Delta\rho(x) 
  = \hbar^{-2}\frac{R}{6}\int_{-\infty}^\infty du (u - s)^3\rho^{(0)\prime}(u) 
    + O(\hbar^{-1}) 
\eeqnn
and
\begin{multline*}
  - \int_{-\infty}^\infty dx 
    \left(\sum_{k=1}^\infty\frac{\hbar^{-1}t_kq^k}{1-q^k}q^{kx}\right)
    \Delta\rho(x) \\
  = - \hbar^{-2}\int_{-\infty}^\infty du 
    \left(\sum_{k=1}^\infty\frac{t_k}{Rk}e^{-Rku}\right)\rho^{(0)\prime}(u)
    + O(\hbar^{-1}).
\end{multline*}
The energy functional thus turns out to be expressed as 
\beq
  \calE^{\mathrm{5D}}_{\hbar^{-1}s,\hbar^{-1}\bst}[\rho]
  = \hbar^{-2}\calE^{\mathrm{5D}(0)}_{s,\bst}[\rho^{(0)}] 
    + O(\hbar^{-1}), 
\eeq
where 
\begin{multline}
  \calE^{\mathrm{5D}(0)}_{s,\bst}[\rho^{(0)}]
  = \iint_{-\infty}^\infty dudvg_{\mathrm{5D}}^{(0)}(|u-v|)
     \rho^{(0)\prime}(u)\rho^{(0)\prime}(v) \\
  \mbox{} 
    + \frac{R}{6}\int_{-\infty}^\infty du(u-s)^3\rho^{(0)\prime}(u) 
    + \int_{-\infty}^\infty du
      \left(\sum_{k=1}^\infty\frac{t_ke^{-Rku}}{-Rk}\right)\rho^{(0)\prime}(u). 
\end{multline}

\subsection{Riemann-Hilbert problem for 5D theory}

Assuming the same ``one-cut'' condition 
(\ref{rho0'-one-cut}) as in the case of the 4D theory, 
one can reduce the minimizing problem 
for $\calE^{\mathrm{5D}(0)}_{s,\bst}[\rho^{(0)}]$ 
to solving the integral equation 
\begin{multline}
  \frac{d}{du} \left( \int_{u_0}^{u_1}dv 
  g_{\mathrm{5D}}^{(0)}(|u-v|)\rho^{(0)\prime}(v) \right)  
    + \frac{R(u-s)^2}{4} + \frac{V(u)}{2} + \frac{\nu}{2} 
  = 0 \\
   (u_0 \le u \le u_1), 
\label{5D-rho0'-eq}
\end{multline}
where 
\beqnn
  V(u) = \sum_{k=1}^\infty t_ke^{-Rku}, 
\eeqnn
under the constraints (\ref{rho0'-constraint}) 
and the support condition 
\beq
  \rho^{(0)\prime}(u) = 0 \quad\text{for $u \not\in [u_0,u_1]$}. 
\eeq

To convert this variational equation to a Riemann-Hilbert problem, 
we introduce the complex analytic function 
\beqnn
  W(z) = \int_{u_0}^{u_1}dv g_{\mathrm{5D}}^{(0)\dprime}(z-v)
         \rho^{(0)\prime}_*(v) 
\eeqnn
with the complexified kernel 
\beqnn
  g_{\mathrm{5D}}^{(0)\dprime}(z-v) 
  = \log\left(\frac{e^{R(z-v)/2} - e^{-R(z-v)/2}}{R\Lambda_0}\right). 
\eeqnn
$W(z)$ is a holomorphic function on the $z$-plane 
with discontinuity along the intervals 
$[u_0+2\pi in/R,\;u_1+2\pi in/R]$, $n \in \ZZ$. 
Moreover, $W(z) + Rz/2$ is periodic with respect 
to the translation $z \mapsto z + 2\pi i/R$.  
The once-differentiated form 
\begin{multline}
  \mathrm{PP} \int_{u_0}^{u_1}dv 
  g_{\mathrm{5D}}^{(0)\dprime}(|u-v|)\rho^{(0)\prime}(v) 
    + \frac{R(u-s)}{2} + \frac{V'(u)}{2} 
  = 0 \\
   (u_0 \le u \le u_1), 
\label{5D-rho0'-eq1}
\end{multline}
of the variational equation (\ref{5D-rho0'-eq}) 
can be converted to the equation 
\beq
  W(u + i0) + W(u - i0) = - R(u - s) - V'(u) 
  \quad (u_0 \le u \le u_1) 
\label{5D-RH(1)}
\eeq
for $W(u\pm i0)$.  This equation is supplemented 
with the equation 
\beq
  W(u + i0) - W(u - i0) 
  = \begin{cases}
    0        &(u > u_1),\\
    - 2\pi i &(u < u_0) 
    \end{cases}
\label{5D-RH(2)}
\eeq
off the support of $\rho^{(0)\prime}_*$. 
The minimizer $\rho^{(0)}_*(u)$ can be read out 
from $W(z)$ by the same formula as (\ref{rho0*-ImW}). 

The constraints (\ref{rho0'-constraint}) 
can be translated to boundary conditions of $W(z)$ 
at infinity.  Unlike the 4D theory, 
$W(z)$ exhibits different boundary behaviors 
as $\Re z \to -\infty$ and $\Re z \to +\infty$: 
\begin{itemize}
\item As $\Re z \to -\infty$ and $\pm \Im z > 0$, 
\beq
  W(z) = \frac{R(z-s)}{2} \mp\pi i + \log(R\Lambda_0) + O(e^{Rz}). 
\label{5D-RH(3)}
\eeq
\item As $\Re z \to +\infty$, 
\beq
  W(z) = - \frac{R(z-s)}{2} + \log(R\Lambda_0) + O(e^{-Rz}).
\label{5D-RH(4)}
\eeq
\end{itemize}
These conditions are derived from (\ref{rho0'-constraint}) 
as follows.  
As $\Re z \to -\infty$, the kernel function behaves as 
\beqnn
  g_{\mathrm{5D}}^{(0)\dprime}(z-v) 
  = - \frac{R(z-v)}{2} \pm\pi i - \log(R\Lambda_0) + O(e^{Rz}). 
\eeqnn
Consequently, as $\Re z \to -\infty$ under 
the condition $\pm\Im z > 0$, $W(z)$ behaves as 
\beqnn
\begin{aligned}
  W(z) 
  &= \int_{u_0}^{u_1}dv \left(- \frac{Rz}{2} + \frac{Rv}{2} 
      \pm \pi i - \log(R\Lambda_0)\right)\rho^{(0)\prime}(v) 
     + O(e^{Rz}) \\
  &= \frac{Rz}{2} - \frac{Rs}{2} \mp\pi i 
     +\log(R\Lambda_0) + O(e^{Rz}).
\end{aligned}
\eeqnn
In much the same way, as $\Re z \to +\infty$, 
\beqnn
  g_{\mathrm{5D}}^{(0)\dprime}(z-v) 
  = \frac{R(z-v)}{2} - \log(R\Lambda_0) + O(e^{-Rz}), 
\eeqnn
hence 
\beqnn
\begin{aligned}
  W(z) 
  &= \int_{u_0}^{u_1}dv \left(\frac{Rz}{2} -\frac{Rv}{2} 
       - \log(R\Lambda_0) \right)\rho^{(0)\prime}(v) 
   + O(e^{-Rz}) \\
  &= - \frac{Rz}{2} + \frac{Rs}{2} + \log(R\Lambda_0) + O(e^{-Rz}). 
\end{aligned}
\eeqnn

Thus $W(z)$ turns out to satisfy (\ref{5D-RH(1)}), 
(\ref{5D-RH(2)}), (\ref{5D-RH(3)}) and (\ref{5D-RH(4)}). 
We have to solve these equations under the condition 
that $W(z) + R(z-s)/2$ be a periodic function 
with respect to the translation $z \mapsto z + 2\pi i/R$.  
In other words, this is a Riemann-Hilbert problem 
on the cylinder $\CC/(2\pi i/R)\ZZ$ with a single cut 
along $[u_0,u_1]$.

\section{Solution of Riemann-Hilbert problem for 4D theory}

\subsection{Construction of solution when $\bst = \bszero$}

When $\bst = \bszero$, we can solve the Riemann-Hilbert problem 
deductively (namely, without any heuristic consideration) 
as follows.  This solution is a prototype of the solution 
in the case of $\bst \not= \bszero$.  

In this case, the equations (\ref{4D-RH(1)}) and (\ref{4D-RH(2)}) 
for $W(u\pm i0)$ read 
\beqnn
  W(u + i0) + W(u - i0) = 0 \quad\text{for $u_0 \le u \le u_1$}
\eeqnn
and
\beqnn
  W(u + i0) - W(u - i0) 
  = \begin{cases} 
     0      &\text{for $u > u_1$},\\
    -2\pi i &\text{for $u < u_0$}. 
    \end{cases}
\eeqnn
One can readily see from these equations 
that the identity 
\beqnn
  e^{W(u+i0)} + e^{-W(u+i0)} = e^{W(u-i0)} + e^{-W(u-i0)} 
\eeqnn
holds along the whole real axis.  This means that 
the complex function $e^{W(z)} + e^{-W(z)}$ has  
no discontinuity, hence becomes a holomorphic function 
on the whole plane $\CC$.  Moreover, by (\ref{4D-RH(3)}), 
this function has the asymptotic form 
\beqnn
  e^{W(z)} + e^{-W(z)} = \frac{z-s}{\Lambda_0} + O(z^{-1}) 
\eeqnn
as $z \to \infty$.  By Liouville's theorem in complex analysis, 
the $O(z^{-1})$ term disappears and one obtains the identity 
\beq
  e^{W(z)} + e^{-W(z)} = \frac{z - s}{\Lambda_0}. 
\eeq
Setting 
\beqnn
  y = e^{-W(z)}, 
\eeqnn
one can rewrite the last identity as 
\beq
  y + y^{-1} = \frac{z - s}{\Lambda_0}. 
\label{4D-SWcurve-t=0}
\eeq

(\ref{4D-SWcurve-t=0}) may be thought of as the equation 
of a complex algebraic curve, which is exactly 
the Seiberg-Witten curve of the undeformed 4D $U(1)$ gauge theory.  
One can solve (\ref{4D-SWcurve-t=0}) for $y = y(z)$ as 
\beq
  y(z) = \frac{z - s + \sqrt{P(z)}}{2\Lambda_0},\quad
  P(z) = (z - s)^2 - 4\Lambda_0^2. 
\eeq
$u_0$ and $u_1$ are determined to be the endpoints 
of the interval of $\RR$ where $P(u) < 0$, namely, 
\beq
  u_0 = s - 2\Lambda_0,\quad u_1 = s + 2\Lambda_0. 
\eeq
$\sqrt{P(z)}$ is understood to be the branch 
on the cut plane $\CC\setminus[u_0,u_1]$ 
such that 
\beq
\begin{gathered}
  \sqrt{P(z)} = z - s + O(z^{-1}) \quad\text{as $z \to \infty$},\\
  \pm\mathrm{Im}\sqrt{P(u\pm i0)} > 0 \quad\text{for $u_0 < u < u_1$}. 
\end{gathered}
\eeq
Thus a (unique) solution of the Riemann-Hilbert problem 
can be obtained explicitly as 
\beq
  W(z) = - \log y(z). 
\eeq

As a cross check, let us examine the behavior 
of $y(z)$ and $W(z)$ along the real axis.  
$y(z)$ has discontinuity along $[u_0,u_1]$, 
and takes real values on $\RR\setminus[u_0,u_1]$. 
The boundary values $y(u\pm i0)$ for $u_0 < u < u_1$ 
are unimodular ($|y(u\pm i0)| = 1$), because 
they are mutually conjugate imaginary solutions 
of (\ref{4D-SWcurve-t=0}).  Bearing these facts in mind, 
one can trace the behavior of $y(u\pm i0)$ 
as $u$ decreases along the real axis.  
When $u$ starts from the right side of $u_1$, 
$y(u)$ decreases towards the value $y(u_1) = 1$.  
When $u$ passes $u_1$, $y(u)$ bifurcates 
into the two imaginary numbers $y(u\pm i0)$, 
which move on the unit circle in opposite directions 
towards $-1$ until $u$ reaches $u_0$.  Accordingly, 
when $u$ decreases from $u = u_1$ to $u = u_0$, 
$\arg y(u \pm i0)$ varies from $0$ to $\pm \pi$.  
For $u < u_0$, $y(u)$ is negative and $\arg y(u\pm i0)$ 
takes the constant values $\pm\pi$.  Translated to 
the language of $W(z)$, this behavior of $\arg y(u\pm i0)$ 
implies exactly that (\ref{4D-RH(2)}) holds. 
By (\ref{rho0*-ImW}), one can read out 
an explicit expression of $\rho^{(0)}_*(u)$: 
\beq
  \rho^{(0)}_*(u)
  = \begin{cases}
    0   & (u > u_1),\\
    \pm \arg y(u\pm i0)/\pi & (u_0 \le u \le u_1),\\
    1   & (u < u_0). 
    \end{cases}
\label{4D-rho0*-t=0}
\eeq
Let us mention that one can further rewrite this result 
to the celebrated ``arc-sine law'' \cite{LS77,VK77}.

\subsection{Construction of solution when $\bst \not= \bszero$}

When $\bst \not= \bszero$, we resort to a heuristic method. 
Namely, we seek a solution of the Riemann-Hilbert problem 
as a deformation of the foregoing solution.  
A naive way will be to add $-V'(z)/2$ as 
\beqnn
  W(z) = - \log y(z) - \frac{V'(z)}{2}, 
\eeqnn
which does satisfy the equations (\ref{4D-RH(1)}) and 
(\ref{4D-RH(2)}) for $W(u\pm i0)$, but not 
the boundary condition (\ref{4D-RH(3)}).  
To fulfill the boundary condition, 
we modify this naive form as 
\beq
  W(z) = - \log y(z) + N(z)\sqrt{P(z)} - \frac{V'(z)}{2}, 
\label{4D-W-formula}
\eeq
where 
\begin{itemize}
\item $y(z)$ is a solution 
\beq
  y(z) = \frac{z-\beta + \sqrt{P(z)}}{2\Lambda},\quad 
  P(z) = (z - \beta)^2 - 4\Lambda^2,   
\eeq
of the equation 
\beq
  y + y^{-1} = \frac{z-\beta}{\Lambda} 
\label{4D-SWcurve}
\eeq
of the deformed Seiberg-Witten curve. 
\item $\beta$ and $\Lambda$ are functions 
$\beta(s,\bst)$ and $\Lambda(s,\bst)$ of 
$s$ and $\bst$ that reduce to the previous values 
at $\bst = \bszero$: 
\beq
  \beta(s,\bszero) = s, \quad 
  \Lambda(s,\bszero) = \Lambda_0
\label{4D-betaLam-t=0}
\eeq
\item $u_0$ and $u_1$ are the zeros of $P(z)$: 
\beq
  u_0 = \beta - 2\Lambda,\quad 
  u_1 = \beta + 2\Lambda. 
\eeq
\item $\sqrt{P(z)}$ is the branch on the cut plane 
$\CC\setminus[u_0,u_1]$ such that 
\beq
\begin{gathered}
  \sqrt{P(z)} = z - \beta + O(z^{-1}) \quad\text{as $z \to \infty$},\\
  \pm\Im\sqrt{P(u\pm i0)} > 0 \quad\text{for $u_0 < u < u_1$}.
\end{gathered}
\eeq
\item $N(z)$ is a linear combination 
\beqnn
  N(z) = \sum_{k=1}^\infty t_kN_k(z) 
\eeqnn
of polynomials $N_k(z)$, $k = 1,2,\cdots$, in $z$.  
\end{itemize}

(\ref{4D-RH(1)}) and (\ref{4D-RH(2)}) are already satisfied 
by the ansatz (\ref{4D-W-formula}). 
To fulfill the boundary condition (\ref{4D-RH(3)}), 
we choose $N_k(z)$ to satisfy the condition 
\beq
  N_k(z) - \frac{kz^{k-1}}{2\sqrt{P(z)}} = O(z^{-1}) 
  \quad \text{as $z \to \infty$}. 
\eeq
$N_k(z)$ is uniquely determined by this condition as 
\beq
  N_1(z) = 0, \quad 
  N_k(z) = \frac{k}{2}(z^{k-2} + c_1z^{k-3} + \cdots + c_{k-2}) 
  \quad\text{for $k \ge 2$}, 
\label{4D-N_k}
\eeq
where $c_1,c_2,\cdots$ are the coefficients of the expansion 
\beqnn
  \frac{1}{\sqrt{P(z)}} = z^{-1} + c_1z^{-2} + c_2z^{-3} + \cdots, 
\eeqnn
and we set $c_0 = 1$ for convenience.  
$N_k(z)$ can be also expressed by using a contour integral as 
\beq
  N_k(z) = \frac{1}{2\pi i}\oint_C \frac{dx}{x-z}
            \frac{kx^{k-1}}{2\sqrt{P(x)}}
           + \frac{kz^{k-1}}{2\sqrt{P(z)}}, 
\eeq
where $C$ is a simple closed curve that encircles 
the interval $[u_0,u_1]$ anti-clockwise 
and leaves $z$ outside.  Rewriting $W(z)$ as 
\beqnn
  W(z) = - \log y(z)
         - \sum_{k=1}^\infty 
           t_k\left(\frac{kz^{k-1}}{2\sqrt{P(z)}} - N_k(z)\right)
           \sqrt{P(z)}, 
\eeqnn
one can see that 
\beqnn
\begin{aligned}
  W(z) 
  &= - \log\frac{z-\beta + \sqrt{P(z)}}{2\Lambda} 
     - \sum_{k=1}^\infty 
       \frac{kt_k}{2}(c_{k-1}z^{-1} + c_kz^{-2} + \cdots)\sqrt{P(z)}\\
  &= - \log\frac{z}{\Lambda} - \sum_{k=1}^\infty \frac{kt_kc_{k-1}}{2} 
     + \left(\beta - \sum_{k=1}^\infty\frac{kt_k(c_k-\beta c_{k-1})}{2}
             \right)z^{-1} 
     + O(z^{-2}) 
\end{aligned}
\eeqnn
as $z \to \infty$.  Matching this expression with (\ref{4D-RH(3)}), 
one obtains the equations 
\beq
  \log\frac{\Lambda}{\Lambda_0} - \sum_{k=1}^\infty \frac{kt_kc_{k-1}}{2} = 0, 
  \label{4D-betaLam-eq1}\\
  \beta - \sum_{k=1}^\infty \frac{kt_k(c_k - \beta c_{k-1})}{2} = s 
  \label{4D-betaLam-eq2}
\eeq
for $\beta = \beta(s,\bst)$ and $\Lambda = \Lambda(s,\bst)$. 

By the implicit function theorem, these equations 
do have a solution in a neighborhood of $\bst = \bszero$ 
that satisfies the initial condition (\ref{4D-betaLam-t=0}). 
Though we omit details, (\ref{4D-betaLam-eq1}) and 
(\ref{4D-betaLam-eq2}) are equivalent to the equations 
that are derived by Marshakov and Nekrasov 
by their method based on the $S$-function (\ref{S-function})
\cite{MN06,Marshakov07}.  

Once $W(z)$ is thus determined, one can use (\ref{rho0*-ImW}) 
to obtain an explicit expression of $\rho^{(0)}_*(u)$: 
\beq
  \rho^{(0)}_*(u) 
  = \begin{cases}
    0  & (u > u_1),\\
    \pm \arg y(u\pm i0)/\pi + N(u)\sqrt{|P(u)|} & (u_0 \le u \le u_1),\\
    1  & (u < u_0). 
    \end{cases}
\label{4D-rho0*}
\eeq
Note that the structure of (\ref{4D-rho0*-t=0}) 
is retained except that a new term proportional 
to $\sqrt{|P(u)|}$ is added.

\subsection{Rewriting solution in terms of Lax function}

By construction, $y = y(z)$ is (a branch of) 
the inverse function of 
\beqnn
  z = z(y) = \beta + \Lambda(y + y^{-1}). 
\eeqnn
We now interpret $z(y)$ to be a long-wave limit 
of the well known Lax operator 
\beq
  \fkL = a(s)e^{\rd_s} + b(s) + a(s-1)e^{-\rd_s}, \quad
  \rd_s = \rd/\rd s, 
\label{1DToda-Lax-op}
\eeq
of the Toda lattice \cite{Flaschka74}, 
where $e^{\pm\rd_s}$ are the shift operators, 
namely, $e^{\pm\rd_s}f(s) = f(s \pm 1)$.  
Note that one can rewrite the last term 
of (\ref{1DToda-Lax-op}) as 
\beq
  a(s-1)e^{-\rd_s} = e^{-\rd_s}\cdot a(s). 
\eeq
In the long-wave (or ``dispersionless'') limit,  
$e^{\rd_s}$ turns into a c-number variable $p$ 
(see Section 6 for details).  We identify $y$ 
with this variable $p$.  

To pursue this analogy further, 
we introduce the truncation notations 
\beqnn
\begin{gathered}
  \left(\sum_{n}a_ny^n\right)_{>0} = \sum_{n>0}a_ny^n, \quad 
  \left(\sum_{n}a_ny^n\right)_{<0} = \sum_{n<0}a_ny^n, \\
  \left(\sum_{n}a_ny^n\right)_m = a_m  \quad (m \in \ZZ)
\end{gathered}
\eeqnn
for Laurent series of $y$;  the same notations 
are used for difference operators as well.  
The time evolutions of the dispersionless 1D Toda hierarchy 
are generated by the Laurent polynomials 
$(z(y)^k)_{>0} - (z(y)^k)_{<0}$, $k = 1,2,\cdots$, 
which are dispersionless limit of the generators 
$(\fkL^k)_{>0} - (\fkL^k)_{<0}$ of time evolutions 
of the 1D Toda hierarchy\cite{Adler79}.  

We now show that the foregoing polynomials $N_k(z)$ 
are closely related to these Laurent polynomials of $y$.  
Note that $(z(y)^k)_{>0} - (z(y)^k)_{<0}$ is a linear combination  
of $y^l - y^{-l}$, $l = 1,2,\cdots$, e.g., 
\begin{align*}
  (z(y))_{>0} - (z(y))_{<0} &= \Lambda(y - y^{-1}),\\
  (z(y)^2)_{>0} - (z(y)^2)_{<0} 
    &= \Lambda^2(y^2-y^{-2}) + 2\Lambda\beta(y-y^{-1}),
       \quad \cdots.
\end{align*}
Since  $y^k - y^{-k}$  can be factorized as 
\beqnn
  y^k - y^{-k}
 = (y^{k-1} + y^{k-2} + \cdots + y^{-k+2} + y^{-k+1})(y - y^{-1}) 
\eeqnn
and $y(z)$ and its inverse satisfy the relations 
\beqnn
  y(z) + y(z)^{-1} = \frac{z - \beta}{\Lambda},\quad 
  y(z) - y(z)^{-1} = \frac{\sqrt{P(z)}}{\Lambda}, 
\eeqnn
$(z(y)^k)_{>0} - (z(y)^k)_{<0}$ can be expressed as 
\beqnn
   (z(y)^k)_{>0} - (z(y)^k)_{<0} = Q_k(z)\sqrt{P(z)}, 
\eeqnn
where $Q_k(z)$ is a polynomial in $z$, e.g., 
\begin{align*}
  (z(y))_{>0} - (z(y))_{<0} &= \sqrt{P(z)},\\
  (z(y)^2)_{>0} - (z(y)^2)_{<0} &= (z + \beta)\sqrt{P(z)}, 
  \quad \cdots.
\end{align*}
The polynomial $Q_k(z)$ can be determined as follows.  
Rewrite the last identity as 
\beqnn
  \frac{(z(y)^k)_{>0} - (z(y)^k)_{<0}}{\sqrt{P(z)}} = Q_k(z)
\eeqnn
and note that 
\beqnn
  \text{LHS} 
  = \frac{z^k}{\sqrt{P(z)}} + O(z^{-1}) 
  \quad \text{as $z \to \infty$}. 
\eeqnn
Truncating the negative powers of $z$ yields the identity 
\beqnn
  Q_k(z) = z^{k-1} + c_1z^{k-2} + \cdots + c_{k-1}. 
\eeqnn
One can thus derive the fundamental formula 
\beq
  (z(y)^k)_{>0} - (z(y)^k)_{<0} 
  =  (z^{k-1} + c_1z^{k-2} + \cdots + c_{k-1})\sqrt{P(z)}. 
\label{4D-(z^k)_{>0,<0}}
\eeq
Comparing this formula with (\ref{4D-N_k}), 
one can readily see that 
\beq
  N_k(z)
  = \frac{k}{2}\frac{(z(y)^{k-1})_{>0} - (z(y)^{k-1})_{<0}}{\sqrt{P(z)}}. 
\eeq
Consequently, the second term on the right hand side 
of (\ref{4D-W-formula}) turns out to be expressed as 
\beq
  N(z)\sqrt{P(z)} 
  = \sum_{k=1}^\infty \frac{kt_k}{2}
    \left( (z(y)^{k-1})_{>0} - (z(y)^{k-1})_{<0}\right). 
\label{4D-N(z)-z(y)}
\eeq

Moreover, since 
\beqnn
  (z(y)^k)_{>0} - (z(y)^k)_{<0} = z^k - (z(y)^k)_0 - 2(z(y)^k)_{<0}, 
\eeqnn
one can rewrite (\ref{4D-(z^k)_{>0,<0}}) as 
\beqnn
\begin{aligned}
  (z(y)^k)_0 + 2(z(y)^k)_{<0} 
  &= z(y)^k - (z(y)^k)_{>0} + (z(y)^k)_{<0}\\
  &= z^k - (z^{k-1} + c_1z^{k-2} +\cdots +c_{k-1})\sqrt{P(z)}\\
  &= (c_kz^{-1} + c_{k+1}z^{-2} + \cdots)\sqrt{P(z)}\\
  &= c_k + (c_{k+1}-\beta c_k)z^{-1} + O(z^{-2}). 
\end{aligned}
\eeqnn
Since $z(y)^{-1} = (\Lambda y)^{-1} + O(y^{-2})$ 
as $y \to \infty$, one can re-expanded the right hand side 
in powers of $y^{-1}$ and pick out the coefficients 
of $y^0$ and $y^{-1}$.  This leads to the identities 
\beq
  (z(y)^k)_0 = c_k, \quad
  2(z(y)^k)_{-1} = \frac{{c}_{k+1}-\beta c_k}{\Lambda}. 
\label{4D-(z^k)_{0,-1}}
\eeq
One can thereby rewrite (\ref{4D-betaLam-eq1}) 
and (\ref{4D-betaLam-eq2}) as 
\begin{gather}
  \log\frac{\Lambda}{\Lambda_0} 
  - \sum_{k=1}^\infty \frac{kt_k}{2}(z(y)^{k-1})_0 
  = 0, \label{4D-betaLam-eq1bis}\\
  \beta 
  - \Lambda \sum_{k=1}^\infty kt_k (z(y)^{k-1})_{-1} 
  = s. \label{4D-betaLam-eq2bis}
\end{gather}
In the next section, we derive the same equations 
from generalized string equations.

\section{Generalized string equations for 4D theory}

\subsection{2D Toda hierarchy}

Let us briefly recall the construction of the Toda hierarchy\cite{UT84} 
\footnote{Since the first paper \cite{UT84} was published, 
the notations for the Toda hierarchy have been considerably changed. 
Our notations are mostly based on the review \cite{TT95}.}
(referred to as the ``2D'' Toda hierarchy to distinguish it 
from the more classical ``1D'' hierarchy\cite{Adler79}). 

This integrable hierarchy is formulated in terms of 
two Lax operators $L$ and $\bar{L}$, which are difference 
(or, so to speak, ``pseudo-difference'') operators 
with respect to the lattice coordinate $s$.  
We now choose the ``symmetric gauge'': 
\beqnn
  L = ae^{\rd_s} + \sum_{n=1}^\infty u_ne^{(1-n)\rd_s},\quad
  \bar{L} = a^{-1}e^{\rd_s} 
     + \sum_{n=1}^\infty\bar{u}_ne^{(1+n)\rd_s}, 
\eeqnn
where $a$, $u_n$'s and $\bar{u}_n$'s 
\footnote{The bar does not mean complex conjugate.}
are dynamical variables that depend on $s$ 
and the two sets of time variables 
$\bsT = (T_1,T_2,\cdots)$ and 
$\bar{\bsT} = (\bar{T}_1,\bar{T}_2,\cdots)$.  
Let us introduce the truncation notations 
\begin{gather*}
  \left(\sum_{n}a_ne^{n\rd_s}\right)_{>0} 
  = \sum_{n>0}a_ne^{n\rd_s}, \quad 
  \left(\sum_{n}a_ne^{n\rd_s}\right)_{<0} 
  = \sum_{n<0}a_ne^{n\rd_s},\\
  \left(\sum_{n}a_ne^{n\rd_s}\right)_m = a_m 
  \quad (m \in \ZZ). 
\end{gather*}
for difference operators.  
  
The time evolutions of the Lax operators are generated 
by Lax equations of the form 
\beq
\begin{gathered}
  \frac{\rd L}{\rd T_k} = [B_k,L],\quad 
  \frac{\rd L}{\rd\bar{T}_k} = [\bar{B}_k,L],\\
  \frac{\rd\bar{L}}{\rd T_k} = [B_k,\bar{L}],\quad 
  \frac{\rd\bar{L}}{\rd\bar{T}_k} = [\bar{B}_k,\bar{L}],
\end{gathered}
\eeq
where 
\beqnn
  B_k = (L^k)_{>0} + \frac{1}{2}(L^k)_0, \quad 
  \bar{B}_k = (\bar{L}^{-k})_{<0} + \frac{1}{2}(\bar{L}^{-k})_0.
\eeqnn
These Lax equations are supplemented by another set 
of Lax equations 
\beq
\begin{gathered}
  \frac{\rd M}{\rd T_k} = [B_k,M],\quad 
  \frac{\rd M}{\rd\bar{T}_k} = [\bar{B}_k,M],\\
  \frac{\rd\bar{M}}{\rd T_k} = [B_k,\bar{M}],\quad 
  \frac{\rd\bar{M}}{\rd\bar{T}_k} = [\bar{B}_k,\bar{M}] 
\end{gathered}
\eeq
and the twisted canonical commutation relations 
\beq
  [L,M] = L,\quad [\bar{L},\bar{M}] = \bar{L} 
\label{Toda-CCR}
\eeq
for the Orlov-Schulman operators 
\beqnn
  M = \sum_{k=1}^\infty kT_kL^k + s + \sum_{n=1}^\infty v_nL^{-n},\quad
  \bar{M} = - \sum_{k=1}^\infty k\bar{T}_k\bar{L}^{-k} + s 
     + \sum_{k=1}^\infty\bar{v}_n\bar{L}^n. 
\eeqnn

\subsection{Dispersionless 2D Toda hierarchy}

The dispersionless Toda hierarchy\cite{TT91} 
is a long-wave limit of the Toda hierarchy.  
In this limit, as briefly mentioned in Section 5, 
the shift operator $e^{\rd_s}$ is replaced by 
a momentum-like variable $p$.  $p$ and $s$ then 
become a (twisted) canonical coordinate system 
of a 2D phase space of classical mechanics. 
This procedure is justified by applying the idea of 
semi-classical (or quasi-classical) approximation 
in quantum mechanics to the Toda hierarchy \cite{TT93}.  

To derive the dispersionless Toda hierarchy, 
one starts from an $\hbar$-dependent formulation 
of the Toda hierarchy, in which $e^{\rd_s}$, $\rd/\rd T_k$ 
and $\rd/\rd\bar{T}_k$ are replaced by $e^{\hbar\rd_s}$, 
$\hbar\rd/\rd T_k$ and $\hbar\rd/\rd\bar{T}_k$.
The Lax and Orlov-Schulman operators are assumed 
to take an $\hbar$-dependent form as 
\begin{gather*}
  L = ae^{\hbar\rd_s} + \sum_{n=1}^\infty u_ne^{(1-n)\hbar\rd_s},\quad
  \bar{L} = a^{-1}e^{\hbar\rd_s} 
     + \sum_{n=1}^\infty\bar{u}_ne^{(1+n)\hbar\rd_s},\\
  M = \sum_{k=1}^\infty kT_kL^k + s + \sum_{n=1}^\infty v_nL^{-n},\quad
  \bar{M} = - \sum_{k=1}^\infty k\bar{T}_k\bar{L}^{-k} + s 
     + \sum_{k=1}^\infty\bar{v}_n\bar{L}^n, 
\end{gather*}
where $a$, $u_n,\bar{u}_n$ and $v_n,\bar{v}_n$ 
depend on $\hbar$ as well as $(s,\bsT,\bar{\bsT})$ 
and have semi-classical expansions 
\begin{gather*}
  a = a^{(0)} + \hbar a^{(1)} + \cdots,\\
  u_n = u_n^{(0)} + \hbar u_n^{(1)} + \cdots,\quad
  \bar{u}_n = \bar{u}_n^{(0)} + \hbar\bar{u}_n^{(1)} + \cdots,\\
  v_n = v_n^{(0)} + \hbar v_n^{(1)} + \cdots,\quad
  \bar{v}_n = \bar{v}_n^{(0)} + \hbar\bar{v}_n^{(1)} + \cdots
\end{gather*}
as $\hbar \to 0$.  They obey the Lax equations
\beq
\begin{gathered}
  \hbar\frac{\rd L}{\rd T_k} = [B_k,L],\quad 
  \hbar\frac{\rd L}{\rd\bar{T}_k} = [\bar{B}_k,L],\\
  \hbar\frac{\rd\bar{L}}{\rd T_k} = [B_k,\bar{L}],\quad 
  \hbar\frac{\rd\bar{L}}{\rd\bar{T}_k} = [\bar{B}_k,\bar{L}],\\
  \cdots\quad (\text{equations of the same form for $M$})\quad\cdots, 
\end{gathered}
\eeq
and the twisted canonical commutation relations 
\beq
  [L,M] = \hbar L, \quad [\bar{L},\bar{M}] = \hbar\bar{L}.
\label{hbarToda-CCR}
\eeq

One can derive, from these equations, the Lax equations 
\beq
\begin{gathered}
  \frac{\rd \calL}{\rd T_k} = \{\calB_k,\calL\},\quad 
  \frac{\rd \calL}{\rd\bar{T}_k} = \{\bar{\calB}_k,\calL\},\\
  \frac{\rd\bar{\calL}}{\rd T_k} = \{\calB_k,\bar{\calL}\},\quad 
  \frac{\rd\bar{\calL}}{\rd\bar{T}_k} = \{\bar{\calB}_k,\bar{\calL}\},\\
  \cdots\quad (\text{equations of the same form for $\calM$})\quad\cdots, 
\end{gathered}
\eeq
and the twisted canonical relations 
\beq
  \{\calL,\calM\} = \calL,\quad \{\bar{\calL},\bar{\calM}\} = \bar{\calL} 
\eeq
for the Laurent series 
\begin{gather*}
  \calL = a^{(0)}p + \sum_{n=1}^\infty u_n^{(0)}p^{1-n},\quad
  \bar{\calL} = a^{(0)-1}p 
     + \sum_{n=1}^\infty\bar{u}_n^{(0)}p^{1+n},\\
  \calM = \sum_{k=1}^\infty kT_k\calL^k + s 
          + \sum_{n=1}^\infty v_n^{(0)}\calL^{-n},\quad
  \bar{\calM} = - \sum_{k=1}^\infty k\bar{T}_k\bar{\calL}^{-k} + s 
     + \sum_{k=1}^\infty\bar{v}_n^{(0)}\bar{\calL}^n 
\end{gather*}
of the new variable $p$.  $\calB_k$ and $\bar{\calB}_k$ are determined 
by $\calL$ and $\bar{\calL}$ as 
\beqnn
  \calB_k = (\calL^k)_{> 0} + \frac{1}{2}(\calL^k)_0,\quad
  \bar{\calB}_k = (\bar{\calL}^{-k})_{<0} + \frac{1}{2}(\bar{\calL}^{-k})_0,
\eeqnn
where $(\quad)_{>0}$, $(\quad)_{<0}$ and $(\quad)_0$ 
are now understood to be truncation operations 
for Laurent series of $p$: 
\begin{gather*}
  \left(\sum_{n}a_np^n\right)_{>0} = \sum_{n>0}a_np^n, \quad 
  \left(\sum_{n}a_np^n\right)_{<0} = \sum_{n<0}a_np^n,\\
  \left(\sum_{n}a_np^n\right)_m = a_m \quad (m \in \ZZ). 
\end{gather*}
$\{\quad,\quad\}$ denotes a Poisson bracket specified below.  
These equations are fundamental building blocks 
of the dispersionless Toda hierarchy.  

This procedure may be thought of as a kind of 
``classical limit'' from quantum mechanics to classical mechanics.  
The Laurent series $\calL,\calM,\bar{\calL},\bar{\calM}$ 
are classical counterparts (referred to as 
the Lax and Orlov-Schulman functions) of $L,M,\bar{L},\bar{M}$, 
in which $e^{\hbar\rd_s}$ turns into the c-number variable $p$, 
and the coefficients are replaced by the leading terms 
of the $\hbar$-expansion.  The Poisson bracket $\{\quad,\quad\}$ 
is defined as 
\beqnn
  \{F,G\} = p\left(\frac{\rd F}{\rd p}\frac{\rd G}{\rd s} 
              - \frac{\rd F}{\rd s}\frac{\rd G}{\rd p}\right) 
\eeqnn
for functions on the 2D phase space $(p,s)$.  
In particular, $p$ and $s$ obey the twisted canonical relation 
\beq
  \{p,s\} = p, 
\eeq
which is exactly the classical limit of the quantum mechanical 
commutation relation 
\beq
  [e^{\hbar\rd_s},s] = \hbar e^{\hbar\rd_s}. 
\eeq

As a technical remark, let us note that 
one can derive the $\hbar$-dependent Toda hierarchy 
from the $\hbar$-independent formulation 
by rescaling the variables $s,\bsT,\bar{\bsT}$ as 
\beq
  s \to \hbar^{-1}s,\quad 
  T_k \to \hbar^{-1}T_k,\quad 
  \bar{T}_k \to \hbar^{-1}\bar{T}_k
\label{rescaling-sTTbar}
\eeq
and simultaneously rescaling $M$ and $\bar{M}$ as 
\footnote{Only after this rescaling, the twisted canonical 
commutation relations of the Lax and Orlov-Schulman operators 
take the correct form (\ref{hbarToda-CCR}).  Actually, 
this ad hoc reasoning can be justified more rigorously 
in the language of an auxiliary linear problem \cite{TT95}.}
\beq
  M \to \hbar^{-1}M, \quad 
  \bar{M} \to \hbar^{-1}\bar{M}. 
\label{rescaling-MMbar}
\eeq

This remark is very important for our purpose.  
Note that (\ref{rescaling-sTTbar}) amounts to 
(\ref{4D-rescaling}) and (\ref{5D-rescaling}) 
in the prescription of thermodynamic limit.
\footnote{(\ref{4D-rescaling}) looks apparently 
different from (\ref{rescaling-MMbar}), but actually, 
they are substantially the same.  A careful inspection 
shows that $H_k$ and $\Phi_k(\mu,s)$ are defined 
to have an excess factor of $\hbar$.  
Because of this factor, $t_k$'s have to be rescaled 
by $\hbar^{-2}$ rather than $\hbar^{-1}$.}
By the way, the energy functionals (\ref{4D-E-functional}) 
and (\ref{5D-E-functional}) depend on $\hbar$ 
(through the relation $q = e^{-R\hbar}$ in the 5D case) 
{\it before} the rescaling.  This is not a contradiction, 
but necessary to obtain a non-trivial thermodynamics limit.  
In a similar sense, to obtain a non-trivial solution 
of the dispersionless Toda hierarchy, one has to start 
from a $\hbar$-{\it dependent} solution 
of the $\hbar$-independent Toda hierarchy, 
and let $\hbar\to 0$  after rescaling the variables 
as (\ref{rescaling-sTTbar}) and (\ref{rescaling-MMbar}).  
Moreover, for the rescaled solution to have a limit 
as $\hbar \to 0$, one has to choose 
the $\hbar$-dependent solution carefully .

\subsection{Generalized string equations}

Since we shall not return to the Toda hierarchy 
in the rest of this paper, let us use 
the same notations $L,M,\bar{L},\bar{M}$, rather than 
$\calL,\calM,\bar{\calL},\bar{\calM}$, 
for the Lax and Orlov-Schulman functions 
of the dispersionless Toda hierarchy.  
They are assumed to be Laurent series of the form 
\begin{gather*}
  L = ap + \sum_{n=1}^\infty u_np^{1-n},\quad
  \bar{L} = a^{-1}p + \sum_{n=1}^\infty\bar{u}_n p^{1+n},\\
  M = \sum_{k=1}^\infty kT_kL^k + s + \sum_{n=1}^\infty v_nL^{-n},\quad
  \bar{M} = - \sum_{k=1}^\infty k\bar{T}_k\bar{L}^{-k} + s 
     + \sum_{k=1}^\infty\bar{v}_n\bar{L}^n. 
\end{gather*}
More precisely, $L$ and $M$ are understood to be 
Laurent series in a neighborhood of $p = \infty$, and 
the other two in a neighborhood of $p = 0$.  
To consider generalized string equations, 
these four Laurent series are further assumed 
to have a common domain of convergence, 
say, $r_0 < |p| < r_1$.
\footnote{One can also consider a genuinely algebraic 
framework in which the coefficients of the Lax and 
Orlov-Schulman functions are meaningful as 
formal Laurent series of $p$.  In such a formulation, 
the coefficients $u_n,v_n,\bar{u}_n,\bar{v}_n$ are treated 
as formal power series of $\bsT$ and $\bar{\bsT}$, 
whose coefficients are functions of $s$ in a suitable class.}

Solutions of the dispersionless Toda hierarchy 
in a {\it general position} \cite{TT95} are characterized 
by functional equations of the form 
\beq
  L = f(\bar{L},\bar{M}), \quad M = g(\bar{L},\bar{M}), 
\label{GSE}
\eeq
where $f = f(z,w)$ and $g = g(z,w)$ are arbitrary functions 
of two variables that satisfy the symplectic condition 
\beq
  \frac{df \wedge dg}{f} = \frac{dz \wedge dw}{z}. 
\label{symp-condition}
\eeq
If the functional equations (\ref{GSE}) 
have a solution, namely a quartet $(L,M,\bar{L},\bar{M})$
of Laurent series of the form shown above, 
they automatically give a solution of the dispersionless 
2D Toda hierarchy \cite{TT95}.  Because of their origin 
in string theories \cite{EK94,Takasaki94,NTT95,KO95,Takasaki96}, 
(\ref{GSE}) are referred to as ``generalized string equations''.  

The goal of our consideration to the end of this section 
is to identify the solution (\ref{4D-W-formula}) 
of the Riemann-Hilbert problem with a solution 
of the particular generalized string equations 
\beq
  L = \bar{L}^{-1},\quad 
  L^{-1}M - \log\frac{L}{a_0} 
  = - \bar{L}\bar{M} + \log\frac{\bar{L}^{-1}}{a_0}, 
\label{4D-GSE1}
\eeq
where $a_0$ is a positive constant that will be 
eventually identified with $\Lambda_0$.  
Let us mention that we have been unable to derive 
(\ref{4D-GSE1}) from the first principle, namely, 
from properties of the partition function. 
We reached (\ref{4D-GSE1}) by guess work seeking 
analogy with the 5D counterparts.  In the case of the 5D theory, 
we know a derivation from the first principle (see Section 7).  

It might be, however, possible to derive (\ref{4D-GSE1}) 
from a suitable modification of the Eguchi-Yang model \cite{EY94}.  
This random matrix model has a logarithmic term 
of the form $\Tr(X\log X - X)$ in the potential, 
which will lead to the logarithmic terms of (\ref{4D-GSE1}).  
See also the work of Nekrasov and Marshakov 
\cite{MN06,Marshakov07} on the origin of logarithmic terms 
in their approach to the thermodynamic limit. 

A few technical remarks on (\ref{4D-GSE1}) will be in order. 
\begin{itemize}
\item (\ref{4D-GSE1}) do not take the form 
shown in (\ref{GSE}).  This is not a serious problem. 
What is crucial is not this standard form itself 
but the symplectic condition (\ref{symp-condition}). 
This condition ensures that the map $(z,w) \mapsto (f(z,w),g(z,w))$ 
connecting $(L,M)$ and $(\bar{L},\bar{M})$ is symplectic 
with respect to the symplectic form $dz/z\wedge dw$.  
In the case of (\ref{4D-GSE1}), 
the relevant map $(z,w) \mapsto (\bar{z},\bar{w})$ 
is defined {\it implicitly} by the equations 
\beq
  z = \bar{z}^{-1},\quad 
  z^{-1}w - \log\frac{z}{a_0} 
  = - \bar{z}\bar{w} + \log\frac{\bar{z}^{-1}}{a_0}. 
\eeq
It is easy to show that this map is indeed symplectic:
\beq
  \frac{dz\wedge dw}{z} = \frac{d\bar{z}\wedge d\bar{w}}{\bar{z}}. 
\eeq
\item The logarithmic terms in (\ref{4D-GSE1}) 
have multi-valuedness in the vicinity of $p = 0$ and $p = \infty$ 
in which the Lax and Orlov-Schulman functions are assumed 
to have Laurent expansion with respect to $p$.  
This problem can be settled by a simple trick 
that we shall present in the course of 
solving (\ref{4D-GSE1}). 
\end{itemize}

\subsection{Solution of generalized string equations}

Let us construct the solution of (\ref{4D-GSE1}) 
that is to be identified with the solution of 
the Riemann-Hilbert problem.  Apart from the trick 
for the logarithmic terms (which resembles 
a trick in the Eguchi-Yang model \cite{EY94}), 
one can proceed in much the same way 
as the case of generalized string equations 
in the large-$N$ Hermitian random matrix model \cite{MAM07}.  

The first equation of (\ref{4D-GSE1}) implies that 
\beqnn
  \fkL := L = \bar{L}^{-1}
\eeqnn
is a Laurent polynomial of the form 
\beq
  \fkL = ap + b + ap^{-1},  
\label{4D-L-formula}
\eeq
where $a$ and $b$ are functions of $s$, $\bsT$ and $\bar{\bsT}$. 
This is nothing but the Lax function of the dispersionless 
1D Toda hierarchy, namely, the long-wave limit of (\ref{1DToda-Lax-op}). 

As regards the second equation of (\ref{4D-GSE1}), 
we add $\log p$ to both hand sides and 
reorganize this equation as 
\beq
  \fkM := L^{-1}M - \log\frac{Lp^{-1}}{a_0} 
       = - \bar{L}\bar{M} + \log\frac{\bar{L}^{-1}p}{a_0}. 
\label{4D-GSE1bis}
\eeq
The modified logarithmic terms, unlike $\log(L/a_0)$ 
and $\log(\bar{L}^{-1}/a_0)$, can be expanded to 
Laurent series of $p$ as 
\beqnn
\begin{aligned}
  \log\frac{Lp^{-1}}{a_0}
  &= \log\frac{a}{a_0} + \log\left(1 + \frac{b}{a}p^{-1} + p^{-2}\right)\\
  &= \log\frac{a}{a_0} + \frac{b}{a}p^{-1} 
     + \left(1 - \frac{b^2}{2a^2}\right)p^{-2} + \cdots 
\end{aligned}
\eeqnn
and 
\beqnn
\begin{aligned}
  \log\frac{\bar{L}^{-1}p}{a_0}
  &= \log\frac{a}{a_0} + \log\left(1 + \frac{b}{a}p + p^2\right)\\
  &= \log\frac{a}{a_0} + \frac{b}{a}p 
     + \left(1 - \frac{b^2}{2a^2}\right)p^2 + \cdots.
\end{aligned}
\eeqnn
In  particular, $\fkM$ itself is a Laurent series of $p$. 

(\ref{4D-GSE1bis}) splits into following two equations: 
\begin{gather}
  \fkM = L^{-1}M - \log\frac{Lp^{-1}}{a_0},
  \label{4D-GSE2}\\
  \fkM = - \bar{L}\bar{M} + \log\frac{\bar{L}^{-1}p}{a_0}.
  \label{4D-GSE3}
\end{gather}
Substituting
\beqnn
  M = \sum_{k=1}^\infty kT_kL^k + s + \sum_{n=1}^\infty v_nL^{-n}, 
\eeqnn
one can rewrite (\ref{4D-GSE2}) as 
\beqnn
  \fkM = \sum_{k=1}^\infty kT_kL^{k-1} + sL^{-1} 
        + \sum_{n=1}^\infty v_nL^{-n-1}
        - \log\frac{a}{a_0} - \frac{b}{a}p^{-1} + \cdots. 
\eeqnn
The $(\quad)_{>0}$ part and the $p^0$ part give the equations 
\begin{gather}
  (\fkM)_{>0} =\sum_{k=2}^\infty kT_k(L^{k-1})_{>0}
             =\sum_{k=2}^\infty kT_k(\fkL^{k-1})_{>0},
  \label{4D-M_{>0}-formula}\\
  (\fkM)_0 = \sum_{k=1}^\infty kT_k(L^{k-1})_0 - \log\frac{a}{a_0}
           = \sum_{k=1}^\infty kT_k(\fkL^{k-1})_0 - \log\frac{a}{a_0}. 
  \label{4D-M_0-formula1}
\end{gather}
In the same way, (\ref{4D-GSE3}) reads 
\beqnn
  \fkM = \sum_{k=1}^\infty k\bar{T}_k\bar{L}^{k+1} - s\bar{L} 
       - \sum_{n=1}^\infty\bar{v}_n\bar{L}^{n+1} 
       + \log\frac{a}{a_0} + \frac{b}{a}p + \cdots, 
\eeqnn
and the $(\quad)_{<0}$ part and the $p^0$ part give the equations 
\begin{gather}
  (\fkM)_{<0} = \sum_{k=2}^\infty k\bar{T}_k(\bar{L}^{1-k})_{<0} 
             = \sum_{k=2}^\infty k\bar{T}_k(\fkL^{k-1})_{<0},
  \label{4D-M_{<0}-formula}\\
  (\fkM)_0   = \sum_{k=1}^\infty k\bar{T}_k(\bar{L}^{1-k})_0 + \log\frac{a}{a_0}
             = \sum_{k=1}^\infty k\bar{T}_k(\fkL^{k-1})_0 + \log\frac{a}{a_0}. 
  \label{4D-M_0-formula2}
\end{gather}
Equating the two expressions (\ref{4D-M_0-formula1}) 
and (\ref{4D-M_0-formula2}) of $(\fkM)_0$, 
one obtains the explicit expression 
\beq
  (\fkM)_0 = \sum_{k=1}^\infty \frac{k(T_k+\bar{T}_k)}{2}(\fkL^{k-1})_0 
\label{4D-M_0-formula} 
\eeq
of $(\fkM)_0$ along with the equation 
\beq
  \log\frac{a}{a_0} 
  = \sum_{k=1}^\infty \frac{k(T_k-\bar{T}_k)}{2}(\fkL^{k-1})_0. 
\label{4D-ab-eq1}
\eeq
From (\ref{4D-M_{>0}-formula}), (\ref{4D-M_{<0}-formula}) 
and (\ref{4D-M_0-formula}), one finds that 
\beq
  \fkM = \sum_{k=2}^\infty kT_k(\fkL^{k-1})_{>0} 
       + \sum_{k=2}^\infty k\bar{T}_k(\fkL^{k-1})_{<0} 
       + \sum_{k=2}^\infty \frac{k(T_k+\bar{T}_k)}{2}(\fkL^{k-1})_0. 
\label{4D-M-formula}
\eeq

Lastly, the $(\quad)_{<0}$ part of (\ref{4D-GSE2}) 
and the $(\quad)_{>0}$ part of (\ref{4D-GSE3}) 
yield some more equations.  The equations 
from the $p^{-1}$ terms of (\ref{4D-GSE2}) 
and those from the $p$ terms of (\ref{4D-GSE3}) 
give the same equations 
\beq
  \frac{s-b}{a} + \sum_{k=2}^\infty k(T_k-\bar{T}_k)(\fkL^{k-1})_{-1} = 0. 
\label{4D-ab-eq2}
\eeq
This equation supplements (\ref{4D-ab-eq1}) 
to determine $a$ and $b$ as functions of $(s,\bsT,\bar{\bsT})$.  
The equations from higher orders of $p^{\pm 1}$ read 
\begin{multline}
  \frac{v_n}{a^{n+1}} 
  + (\text{linear combination of $v_1,\cdots,v_{n-1}$})\\
  = - \sum_{k=1}^\infty k(T_k-\bar{T}_k)(\fkL^{k-1})_{-n-1}
    + \left(\log(Lp^{-1}/a_0)\right)_{-n-1},
\end{multline}
and 
\begin{multline}
  \frac{\bar{v}_n}{a^{n+1}}
  + (\text{linear combination of $\bar{v}_1,\cdots,\bar{v}_1$})\\
  = - \sum_{k=1}^\infty k(T_k-\bar{T}_k)(\fkL^{k-1})_{n+1} 
    + \left(\log(\bar{L}^{-1}p/a_0)\right)_{n+1}. 
\end{multline}
These equations determine $v_n$ and $\bar{v}_n$ recursively 
once $a$ and $b$ are obtained.  One can thus construct 
a (unique) solution of the generalized string equations 
(\ref{4D-GSE1}).

\subsection{Identification}

As remarked above, the first equation of (\ref{4D-GSE1}) 
is a reduction condition to the dispersionless 1D Toda hierarchy. 
If this equation holds, then 
\beq
  B_k + \bar{B}_k = \fkL^k, 
\eeq
hence 
\beq
  \frac{\rd\fkL}{\rd T_k} + \frac{\rd\fkL}{\rd\bar{T}_k} 
  = \{\fkL^k, \fkL\} 
  = 0. 
\eeq
This implies that $a$ and $b$ (and, actually, all other 
dynamical variables) depend on $\bsT$ and $\bar{\bsT}$ 
through the difference $\bsT-\bar{\bsT}$ only: 
\beq
  a = a(s, \bsT-\bar{\bsT}), \quad 
  b = b(s, \bsT-\bar{\bsT}).
\eeq
This explains why $T_k$ and $\bar{T}_k$ 
show up in (\ref{4D-ab-eq1}) and (\ref{4D-ab-eq2}) 
in the form of the difference $T_k - \bar{T}_k$. 

Let us now restrict the time evolutions 
to the anti-diagonal subspace 
\beq
  T_k = - \bar{T}_k = \frac{t_k}{2}, \quad k = 1,2,\cdots. 
\label{4D-T=-Tbar}
\eeq
The reduced Lax function $\fkL$ therein satisfies 
the standard Lax equations 
\beq
  \frac{\rd\fkL}{\rd t_k} = \{A_k,\fkL\}, \quad 
  A_k =\frac{1}{2}(\fkL^k)_{>0} - \frac{1}{2}(\fkL^k)_{<0} 
\eeq
of the dispersionless 1D Toda hierarchy.  
Moreover, (\ref{4D-ab-eq1}), (\ref{4D-ab-eq2}) 
and  (\ref{4D-M-formula}) take the reduced form 
\begin{gather}
  \log\frac{a}{a_0} = \sum_{k=1}^\infty \frac{kt_k}{2}(\fkL^{k-1})_0,
  \label{4D-ab-eq1bis}\\
  \frac{s-b}{a} + \sum_{k=2}^\infty kt_k(\fkL^{k-1})_{-1} = 0,
  \label{4D-ab-eq2bis}\\
  \fkM = \sum_{k=2}^\infty \frac{kt_k}{2}
         \left((\fkL^{k-1})_{>0} - (\fkL^{k-1})_{<0}\right). 
  \label{4D-M-formula1}
\end{gather}
For (\ref{4D-ab-eq1bis}) and (\ref{4D-ab-eq2bis}) to coincide 
with (\ref{4D-betaLam-eq1}) and (\ref{4D-betaLam-eq2}), 
it is sufficient that the building blocks 
of the Riemann-Hilbert problem and the generalized string equations 
are identified as 
\beq
  a_0 = \Lambda_0, \quad a = \Lambda, \quad b = \beta,\quad 
  y = p, \quad z(y) = \fkL. 
\label{4D-RH-vs-dToda}
\eeq
Comparing (\ref{4D-M-formula1}) with (\ref{4D-N(z)-z(y)}), 
one finds that that $W(z)$ and $\fkM$ are related as 
\beq
  W(z) + \log y(z) + \frac{V'(z)}{2} = N(z)\sqrt{P(z)} = \fkM.  
\eeq

Thus (\ref{4D-W-formula}) can be identified with 
a solution of the dispersionless Toda hierarchy 
that satisfies the generalized string equations (\ref{4D-GSE1}). 
Let us stress that the generalized string equations 
are a nonlinear analogue of (so to speak, {\it modern}) 
matrix Riemann-Hilbert problems that are widely used 
in the theory of integrable systems \cite{TT95}. 
We thus have a nonlinear ({\it ultra-modern}) 
Riemann-Hilbert problem for $L,M,\bar{L},\bar{M}$ on the $p$-plane. 
On the other hand, we have another ({\it classic}) 
Riemann-Hilbert problem for $W(z)$ on the $z$-plane. 
This is a very interesting interplay of 
two Riemann-Hilbert problems of quite different natures.

\section{Solution of Riemann-Hilbert problem for 5D theory}

\subsection{Construction of solution when $\bst = \bszero$}

As in the case of the 4D theory, we can deductively solve 
the Riemann-Hilbert problem in the case where  $\bst = \bszero$. 

When $\bst = \bszero$, one can rewrite (\ref{5D-RH(1)}) 
and (\ref{5D-RH(2)}) as 
\beqnn
  \left(W(u+i0) + \frac{R(u-s)}{2}\right)
  + \left(W(u-i0) + \frac{R(u-s)}{2}\right) 
  = 0 
  \quad \text{for $u_0 \le u \le u_1$} 
\eeqnn
and 
\beqnn
  \left(W(u+i0) + \frac{R(u-s)}{2}\right)
  - \left(W(u-i0) + \frac{R(u-s)}{2}\right) 
  = \begin{cases}
    0        & (u > u_1),\\
    - 2\pi i & (u < u_0).
    \end{cases}
\eeqnn
One can see from these equations that the complex function 
$e^{W(z)+R(z-s)/2} + e^{-W(z)-R(z-s)/2}$ has no discontinuity 
along the real axis, hence becomes a holomorphic function 
on the whole $z$-plane $\CC$.  Moreover, 
by the other conditions in the Riemann-Hilbert problem, 
this function is periodic with respect to 
the translation $z \mapsto z + 2\pi i/R$ 
and behaves as 
\beqnn
  e^{W(z)+R(z-s)/2} + e^{-W(z)-R(z-s)/2}
  = - \frac{e^{-R(z-s)}}{R\Lambda_0} + O(1) 
\eeqnn
as $\Re z \to -\infty$ and 
\beqnn
  e^{W(z)+R(z-s)/2} + e^{-W(z)-R(z-s)/2}
  = \frac{1}{R\Lambda_0} + R\Lambda_0 + O(e^{-Rz})  
\eeqnn
as $\Re z \to +\infty$.  Therefore, by Liouville's theorem, 
one can deduce that the identity 
\beq
  e^{W(z)+R(z-s)/2} + e^{-W(z)-R(z-s)/2} 
  = \frac{1}{R\Lambda_0} + R\Lambda_0 - \frac{e^{-R(z-s)}}{R\Lambda_0} 
\eeq
holds.  

Let us introduce 
\beqnn
  y = e^{-W(z)-R(z-s)/2}, \quad 
  \beta = (1 + (R\Lambda_0)^2)e^{-Rs},\quad 
  \Lambda = \Lambda_0 e^{-Rs}
\eeqnn
to rewrite the last identity as 
\beq
  y + y^{-1} = \frac{\beta - e^{-Rz}}{R\Lambda}. 
\label{5D-SWcurve-t=0}
\eeq
Viewed as an equation of a complex analytic curve, 
(\ref{5D-SWcurve-t=0}) defines substantially 
the same Seiberg-Witten curve as previously considered 
for the undeformed 5D $U(1)$ gauge theory \cite{Maeda-Nakatsu06}. 

One can solve this equation for $y$ as 
\beq
  y = y(z) = \frac{\beta - e^{-Rz} - \sqrt{P(z)}}{2R\Lambda},\quad
  P(z) = (e^{-Rz} - \beta)^2 - 4(R\Lambda)^2. 
\eeq
$u_0$ and $u_1$ are determined to be the endpoint 
of the interval of $\RR$ where $P(u) < 0$, namely, 
\beq
\begin{gathered}
  u_0 = - \frac{\log(\beta + 2R\Lambda)}{R}
      = s - \frac{2\log(1 + R\Lambda_0)}{R},\\
  u_1 = - \frac{\log(\beta - 2R\Lambda)}{R} 
      = s - \frac{2\log(1 - R\Lambda_0)}{R}. 
\end{gathered}
\eeq
$\sqrt{P(z)}$ is the branch on the $z$-plane 
cut along the intervals 
$[u_0 + 2\pi in/R,\,u_1 + 2\pi in/R$, $n \in \ZZ$, 
such that 
\beq
\begin{gathered}
  \sqrt{P(z)} = e^{-Rz} - \beta + O(e^{Rz}) 
    \quad \text{as $\Re z \to -\infty$},\\
  \sqrt{P(z)} = - \sqrt{\beta^2 - 4(R\Lambda)^2} + O(e^{-Rz}) 
    \quad \text{as $\Re z \to +\infty$},\\
  \mp\mathrm{Im}\sqrt{P(u\pm i0)} > 0 \quad \text{for $u_0 < u < u_1$}. 
\end{gathered}
\eeq

Solving 
\beqnn
  e^{-W(z)-R(z-s)/2} = y(z)
\eeqnn
for $W(z)$ yields a (unique) solution of 
the Riemann-Hilbert problem explicitly as 
\beq
  W(z) = - \frac{R(z-s)}{2} - \log y(z). 
\eeq
As explained in the case of the 4D theory, 
it is easy to reconfirm directly that this function 
fulfills all requirements of the Riemann-Hilbert problems.  
The minimizer $\rho^{(0)}_*(u)$ has the same expression 
as the 4D case (\ref{4D-rho0*-t=0}).

\subsection{Construction of solution when $\bst \not= \bszero$}

Let us now consider the case where $\bst \not= \bszero$.  
We seek a solution of the Riemann-Hilbert problem 
by modifying the solution for $\bst \not= \bszero$ as 
\beq
  W(z) =  - \frac{R}{2}(z-s) - \log y(z) 
          + N(z)\sqrt{P(z)} - \frac{V'(z)}{2}, 
\label{5D-W-formula}
\eeq
where 
\begin{itemize}
\item $y(z)$ is a solution 
\beq
  y(z) = \frac{\beta - e^{-Rz} - \sqrt{P(z)}}{2R\Lambda},\quad 
  P(z) = (e^{-Rz}-\beta)^2 - 4(R\Lambda)^2, 
\eeq
of the equation 
\beq
  y + y^{-1} = \frac{\beta - e^{-Rz}}{R\Lambda} 
\label{5D-SWcurve}
\eeq
of the deformed Seiberg-Witten curve.  
\item $\beta$ and $\Lambda$ are functions $\beta(s,\bst)$ 
and $\Lambda(s,\bst)$ of $s$ and $\bst$ that reduces 
to the previous values at $\bst = \bszero$: 
\beq
  \beta(s,\bszero) = (1 + (R\Lambda_0)^2)e^{-Rs},\quad 
  \Lambda(s,\bszero) = \Lambda_0e^{-Rs}. 
\label{5D-betaLam-t=0}
\eeq
\item $u_0$ and $u_1$ are the real zeros of $P(z)$: 
\beq
  u_0 = - \frac{\log(\beta + 2R\Lambda)}{R},\quad 
  u_1 = - \frac{\log(\beta - 2R\Lambda)}{R}. 
\eeq
\item $\sqrt{P(z)}$ is the branch on the $z$-plane 
cut along the intervals $[u_0+2\pi in/R, u_1+2\pi in/R]$, 
$n \in \ZZ$, such that 
\beq
\begin{gathered}
  \sqrt{P(z)} = e^{-Rz} - \beta + O(e^{Rz}) 
    \quad \text{as $\Re z \to -\infty$},\\
  \sqrt{P(z)} = - \sqrt{\beta^2 - 4(R\Lambda)^2} + O(e^{-Rz}) 
    \quad \text{as $\Re z \to +\infty$},\\
  \mp\Im\sqrt{P(u\pm i0)} > 0 \quad \text{for $u_0 < u < u_1$}. 
\end{gathered}
\eeq
\item $N(z)$ is a linear combination 
\beqnn
  N(z) = \sum_{k=1}^\infty t_kN_k(z) 
\eeqnn
of polynomials $N_k(z)$ in $e^{-Rz}$. 
\end{itemize}

(\ref{5D-RH(1)}) and (\ref{5D-RH(2)}) are already satisfied 
by the ansatz (\ref{5D-W-formula}).  Moreover, 
since $P(z)$, $N(z)$ and $V(z)$ are periodic 
with respect to the translation $z \mapsto z + 2\pi i/R$, 
the periodicity condition of $W(z) + R(z-s)/2$  
is also satisfied. Therefore it is sufficient to fulfill 
the boundary conditions (\ref{5D-RH(3)}) and (\ref{5D-RH(4)}). 

To this end, we choose $N_k(z)$ to satisfy the condition 
\beq
  N_k(z) + \frac{Rke^{-Rkz}}{2\sqrt{P(z)}} = O(e^{Rz}) 
  \quad \text{as $\Re z \to -\infty$}. 
\eeq
$N_k(z)$ is uniquely determined by this condition as 
\beq
  N_k(z) = - \frac{Rk}{2}(e^{-R(k-1)z} + c_1e^{-R(k-2)z} 
           + \cdots + c_{k-1}), 
\eeq
where $c_1,c_2,\cdots$ are the coefficients in the expansion 
\beqnn
  \frac{1}{\sqrt{P(z)}} = e^{Rz} + c_1e^{2Rz} + c_2e^{3Rz}+ \cdots, 
\eeqnn
and we set $c_0 = 1$ for convenience.  
Consequently, as $\Re z \to -\infty$ in the domain $\pm\Im z > 0$, 
$W(z)$ behaves as 
\beqnn
\begin{aligned}
  W(z) 
  &= - \frac{R(z-s)}{2} 
          - \log\frac{\beta-e^{-Rz}-\sqrt{P(z)}}{2R\Lambda}\\
  &\quad + \sum_{k=1}^\infty \frac{Rkt_k}{2}
           \left(c_ke^{Rz} + c_{k+1}e^{R(k+1)z} +\cdots
           \right)\sqrt{P(z)}\\
  &= \frac{R(z+s)}{2}  \mp\pi i + \log(R\Lambda)  
     + \sum_{k=1}^\infty \frac{Rkt_kc_{k}}{2} + O(e^{Rz}). 
\end{aligned}
\eeqnn
One thus finds the equation 
\beq
  Rs + \log\frac{\Lambda}{\Lambda_0} 
  + \sum_{k=1}^\infty \frac{Rkt_kc_k}{2} 
  = 0 
\label{5D-betaLam-eq1}
\eeq
for the first boundary condition (\ref{5D-RH(3)}) 
to be satisfied.  

On the other hand, in the opposite end 
($\Re z \to +\infty$) of the $z$-plane, 
$N(z)$, $P(z)$ and $y(z)$ converge to finite values as 
\begin{gather*}
  \lim_{\Re z\to+\infty}N(z) = - \sum_{k=1}^\infty \frac{Rkt_kc_{k-1}}{2},\\
  \lim_{\Re z\to+\infty}\sqrt{P(z)} = - \sqrt{\beta^2 - 4(R\Lambda)^2},\\
  y_\infty: = \lim_{\Re z\to+\infty}y(z) 
    =  \frac{\beta + \sqrt{\beta^2-4(R\Lambda)^2}}{2R\Lambda} 
\end{gather*}
and $V(z)$ disappears.  Therefore $W(z)$ behaves as 
\beqnn
  W(z) 
  = - \frac{R(z-s)}{2} - \log y_\infty 
    + \sum_{k=1}^\infty\frac{Rkt_kc_{k-1}}{2}\sqrt{\beta^2 - 4(R\Lambda)^2} 
    + O(e^{-Rz}), 
\eeqnn
One thus obtains the equation 
\beq
  - \log y_\infty
  + \sum_{k=1}^\infty \frac{Rkt_kc_{k-1}}{2}\sqrt{\beta^2 - 4(R\Lambda)^2} 
  = \log(R\Lambda_0) 
\label{5D-betaLam-eq2}
\eeq
for the second boundary condition (\ref{5D-RH(4)}) 
to be satisfied. 

Thus the problem reduces to solving 
(\ref{5D-betaLam-eq1}) and (\ref{5D-betaLam-eq2}) 
for $\beta = \beta(s,\bst)$ and $\Lambda = \Lambda(s,\bst)$.  
The implicit function theorem ensures 
the existence of a solution in a neighborhood 
of $\bst = \bszero$ that satisfy the initial conditions 
(\ref{5D-betaLam-t=0}).  Note that (\ref{5D-betaLam-eq1}) 
and (\ref{5D-betaLam-eq2}) reduce to 
\beqnn
  Rs + \log\frac{\Lambda}{\Lambda_0} = 0, \quad 
  \frac{\beta+\sqrt{\beta^2-4(R\Lambda)^2}}{2R\Lambda} 
  =\frac{1}{R\Lambda_0} 
\eeqnn
when $\bst = \bszero$.  It is easy to see that 
(\ref{5D-betaLam-t=0}) give a solution of these equations. 
Lastly, the the minimizer $\rho^{(0)}_*(u)$ turns out 
to have the same expression as the 4D case (\ref{4D-rho0*}).

\subsection{Rewriting solution in terms of Lax function}

Let us  rewrite the foregoing solution 
of the Riemann-Hilbert problem in terms of the new variables 
\beqnn
  Z = e^{-Rz}. 
\eeqnn
Note that the cylinder $\CC/(2\pi i/R)\ZZ$ is thereby mapped 
to the punctured plane $\CC^* = \CC\setminus\{0\}$.  
$P(z)$ and $N_k(z)$ thereby turn into polynomials in $Z$ as 
\beq
\begin{gathered}
  P(z) = (Z-\beta)^2 - 4(R\Lambda)^2, \\
  N_k(z) = - \frac{Rk}{2}(Z^{k-1} + c_1Z^{k-2} + \cdots + c_{k-1}). 
\end{gathered}
\eeq
$c_n$'s may be thought of as the coefficients of 
the Laurent expansion 
\beqnn
  \frac{1}{\sqrt{(Z-\beta)^2-4(R\Lambda)^2}}
  = Z^{-1} + c_2Z^{-2} + c_3Z^{-3} + \cdots 
\eeqnn
at $Z = \infty$, and $N_k(z)$'s are uniquely determined 
by the condition 
\beqnn
  N_k(z) + \frac{RkZ^k}{2\sqrt{(Z-\beta)^2-4(R\Lambda)^2}} = O(Z^{-1}) 
  \quad \text{as $Z \to \infty$}. 
\eeqnn
Another expression of $N_k(z)$'s is the contour integral formula 
\beq
  N_k(z) = - \frac{1}{2\pi i}\oint_C\frac{dX}{X-Z}
             \frac{RkX^k}{2\sqrt{(X-\beta)^2-4(R\Lambda)^2}}
           - \frac{RkZ^{k}}{2\sqrt{(Z-\beta)^2-4(R\Lambda)^2}}, 
\nonumber \\           
\eeq
where $C$ is a simple closed curve that encircles 
the interval $[e^{-Ru_1},e^{-Ru_0}]$ anti-clockwise 
and leaves $e^{-Rz}$ outside.  

The function $y(z)$ can be redefined as a function 
of $Z$ as 
\beq
  y(Z) = \frac{\beta - Z - \sqrt{P(z)}}{2R\Lambda}, 
\eeq
which is an inverse function of 
\beqnn
  Z(y) = \beta - R\Lambda(y + y^{-1}).  
\eeqnn
This function $Z(y)$ plays the role of a Lax function. 
\footnote{Strictly speaking, it is $-Z(y)$ rather than $Z(y)$ 
that correspond to the Lax function $\fkL$ (see section 8).}
As in the case of the 4D theory, one can derive 
the following identities: 
\begin{gather}
  (Z(y)^k)_{>0} - (Z(y)^k)_{<0} 
  = (Z^{k-1} + c_1Z^{k-2} + \cdots + c_{k-1})\sqrt{P(z)}, \\
  (Z(y)^k)_0 = c_k, \quad 
  (Z(y)^k)_{-1} = -\frac{c_{k+1}-\beta c_k}{2R\Lambda}. 
\end{gather}
One can thereby rewrite the third term of (\ref{5D-W-formula}) as 
\beq
  N(z)\sqrt{P(z)} 
  = - \sum_{k=1}^\infty\frac{Rkt_k}{2}
      \left((Z(y)^k)_{>0} - (Z(y)^k)_{<0}\right) 
\label{5D-N(z)-Z(y)}
\eeq
and (\ref{5D-betaLam-eq1}) as 
\beq
  Rs + \log\frac{\Lambda}{\Lambda_0} 
  + \sum_{k=1}^\infty \frac{Rkt_k}{2}(Z(y)^k)_0
  = 0. 
\label{5D-betaLam-eq1bis}
\eeq

Unlike (\ref{5D-RH(1)}), (\ref{5D-RH(2)}) and 
(\ref{5D-RH(3)}), it seems difficult to interpret 
(\ref{5D-betaLam-eq2}) correctly within this framework.   
(\ref{5D-betaLam-eq2}) is derived to fulfill 
the boundary condition (\ref{5D-RH(4)}) as 
$\Re z \to +\infty$, in other words, $Z \to 0$ 
(rather than $Z \to \infty$).  The situation of the 5D theory 
thus turns out to be considerably different from the 4D theory.

\section{Solution of dispersionless Toda hierarchy for 5D theory}

\subsection{Generalized string equations}

The goal of this section is two-fold.  Firstly, 
following the method of Section 6, 
we show that the solution (\ref{5D-W-formula}) 
of the Riemann-Hilbert problem corresponds to a solution 
of the particular generalized string equations 
\beq
  L = \bar{L}^{-1},\quad 
  e^{-RM} = Q^{-1}\bar{L}^{-2}e^{R\bar{M}}. 
\label{5D-GSE}
\eeq
Secondly, we derive these equations from hidden symmetries 
of a tau function $\tau_s(\bsT,\bar{\bsT})$ 
of the Toda hierarchy \cite{NT07,NT08}.  
As we briefly review therein, this tau function 
reduces to the partition function $Z^{\mathrm{5D}}_s(\bst)$ 
when the time variables $(\bsT,\bar{\bsT})$ 
are restricted to the anti-diagonal subspace. 

Let us remark that (\ref{5D-GSE}) can be converted 
to a form that resembles the generalized string equations 
(\ref{4D-GSE1}) of the 4D theory.  
Rewrite the term $Q^{-1}\bar{L}^{-2}$ 
on the right hand side of the second equation as 
\beqnn
  Q^{-1}\bar{L}^{-2} 
  = \frac{L}{Q^{1/2}}\cdot\frac{\bar{L}^{-1}}{Q^{1/2}}, 
\eeqnn
move the factor $L/Q^{-1/2}$ to the left hand side 
and take the logarithm of both hand sides.  
One thus obtains the equations 
\beq
  L = \bar{L}^{-1}, \quad 
  - RM - \log\frac{L}{Q^{1/2}} 
    = R\bar{M} + \log\frac{\bar{L}^{-1}}{Q^{1/2}} 
\label{5D-GSE1}
\eeq
with remarkable similarity with (\ref{4D-GSE1}). 

Unfortunately, (\ref{5D-GSE1}) turn out to be 
not strong enough to characterize the solution 
of the Riemann-Hilbert problem uniquely. As one finds 
in the course of solving these string equations, 
(\ref{5D-RH(1)}), (\ref{5D-RH(2)}) and (\ref{5D-RH(3)}) 
are correctly encoded in the generalized string equations, 
but the second boundary condition (\ref{5D-RH(4)}) 
is missing therein.  Because of this, 
our solution of the generalized sting equations 
contains arbitrariness.  This arbitrariness 
can be fixed by imposing (\ref{5D-RH(4)}) by hand, 
but its status in the framework of the dispersionless 
Toda hierarchy is obscure.  

Thus, although looking very similar, 
the 5D generalized string equations (\ref{5D-GSE1}) 
are drastically different from the 4D version 
(\ref{4D-GSE1}).  We can point out technical reasons 
leading to this difference, but a true reason 
is beyond our scope.

\subsection{Solving generalized string equations}

The method for solving (\ref{5D-GSE1}) 
is parallel to the case of the 4D theory.  

The first equation of (\ref{5D-GSE1}) 
again leads to the Lax function 
\beq
  \fkL := L = \bar{L}^{-1} = ap + b + ap^{-1}
\eeq
of the dispersionless 1D Toda hierarchy.  
$a = a(s,\bsT,\bar{\bsT})$ and 
$b = b(s,\bsT,\bar{\bsT})$ are arbitrary at this stage.  
We now derive equations for these functions 
from the the second equation of (\ref{5D-GSE1}).  

Adding $\log p$ to both hand sides, one can rewrite 
the second equation of (\ref{5D-GSE1}) as 
\beq
  \fkM := - RM - \log\frac{Lp^{-1}}{Q^{1/2}} 
        = R\bar{M} + \log\frac{\bar{L}^{-1}p}{Q^{1/2}}, 
\eeq
and separate this this equation to two equations: 
\begin{gather}
  \fkM = - RM - \log\frac{Lp^{-1}}{Q^{1/2}}, 
  \label{5D-GSE2}\\
  \fkM = R\bar{M} + \log\frac{\bar{L}^{-1}p}{Q^{1/2}}. 
  \label{5D-GSE3}
\end{gather}
The $(\quad)_{>0}$ parts of (\ref{5D-GSE2}) 
and the $(\quad)_{<0}$ part of  (\ref{5D-GSE3}) give 
\beq
  (\fkM)_{>0} = - \sum_{k=1}^\infty RkT_k(\fkL^k)_{>0},\quad
  (\fkM)_{<0} = - \sum_{k=1}^\infty Rk\bar{T}_k(\fkL^k)_{<0}. 
\eeq
The $(\quad)_0$ parts of (\ref{5D-GSE2}) 
and (\ref{5D-GSE3}) read 
\begin{gather}
  (\fkM)_0 = - Rs - \log\frac{a}{Q^{1/2}} 
              - \sum_{k=1}^\infty RkT_k(\fkL^k)_0,\\
  (\fkM)_0 = Rs + \log\frac{a}{Q^{1/2}} 
             - \sum_{k=1}^\infty Rk\bar{T}_k(\fkL^k)_0.
\end{gather}
Equating these two expressions yields the explicit expression 
\beq
  (\fkM)_0 = - \sum_{k=1}^\infty \frac{Rk(T_k + \bar{T}_k)}{2}(\fkL^k)_0 
\eeq
of $(\fkM)_0$ and the equation 
\beq
  Rs + \log\frac{a}{Q^{1/2}}
  + \sum_{k=1}^\infty \frac{Rk(T_k-\bar{T}_k)}{2}(\fkL^k)_0 
  = 0. 
  \label{5D-ab-eq1}
\eeq
Note that (\ref{5D-ab-eq1}) amounts to (\ref{4D-ab-eq1}) 
in the case of 4D theory. Thus one finds that 
\beq
  \fkM = - \sum_{k=1}^\infty RkT_k(\fkL^k)_{>0} 
       - \sum_{k=1}^\infty Rk\bar{T}_k(\fkL^k)_{<0} 
       - \sum_{k=1}^\infty \frac{Rk(T_k+\bar{T}_k)}{2}(\fkL^k)_0. 
\label{5D-M-formula}
\eeq

On the other hand, from the $(\quad)_{<0}$ part of 
(\ref{5D-GSE2}) and the $(\quad)_{>0}$ part of 
(\ref{5D-GSE3}), one obtains the recursive formulae 
\begin{multline}
  \frac{Rv_n}{a^n} 
  + (\text{linear combination of $v_1,\cdots,v_{n-1}$})\\
  = - \sum_{k=1}^\infty Rk(T_k - \bar{T}_k)(\fkL^k)_{-n}
    - \left(\log(Lp^{-1}/Q^{1/2})\right)_{-n}
\end{multline}
and 
\begin{multline}
  \frac{R\bar{v}_n}{a^n} 
  + (\text{linear combination of $\bar{v}_1,\cdots,\bar{v}_1$})\\
  = - \sum_{k=1}^\infty Rk(T_k - \bar{T}_k)(\fkL^k)_n 
    - \left(\log(\bar{L}^{-1}p/Q^{1/2})\right)_n
\end{multline}
of $v_n$ and $\bar{v}_n$ for $n = 1,2,\cdots$.  
Note that these equations are all that one can derive 
from the $(\quad)_{<0}$ part of (\ref{5D-GSE2}) 
and the $(\quad)_{>0}$ part of (\ref{5D-GSE3}).  
There is no equation that amounts to (\ref{4D-ab-eq2}) 
in the case of the 4D theory.  

Thus, unlike the case of the 4D theory, 
we have only one equation (\ref{5D-ab-eq1}) 
for the two unknown functions $a$ and $b$. 
Therefore these functions remain to be fully determined.  
This is the aforementioned ``arbitrariness''. 
In this sense, these generalized string equations 
(\ref{5D-GSE1}) are incomplete.  

From a genuinely technical point of view, 
this phenomenon stems from a delicate difference 
of the structures of (\ref{4D-GSE1}) and (\ref{5D-GSE1}). 
In the second equation of (\ref{4D-GSE1}), 
$M$ and $\bar{M}$ are multiplied by $L^{-1}$ and $\bar{L}$.  
Because of the presence of these multipliers, 
when this equation is split into (\ref{4D-GSE2}) 
and (\ref{4D-GSE3}) and expanded in powers of $p$, 
the $p^{-1}$ terms in (\ref{4D-GSE2}) and 
the $p$ terms in (\ref{4D-GSE3}) yield exceptional equations.  
These equations, which are mutually equivalent, 
are nothing but the second equation (\ref{4D-ab-eq2}) 
for $a$ and $b$.  In the case of the second equation 
of (\ref{5D-GSE1}), there is no such exceptional terms, 
hence no counterpart of (\ref{4D-ab-eq2}).  

Let us now restrict the time evolutions to 
the anti-diagonal subspace.  As the main result 
of our previous work \cite{NT07} predicts, 
the reduced time variables $t_k$ should be twisted 
by the signature factor $(-1)^k$ as 
\beq
  T_k = - \bar{T}_k = (-1)^k\frac{t_k}{2}, \quad 
  k = 1,2,\cdots. 
\label{5D-T=-Tbar}
\eeq
(\ref{5D-ab-eq1}) and (\ref{5D-M-formula}) then 
become the following equations: 
\begin{gather}
  Rs + \log\frac{a}{Q^{1/2}}
  + \sum_{k=1}^\infty \frac{(-1)^kRkt_k}{2}(\fkL^k)_0 
  = 0, \label{5D-ab-eq1bis}\\
  \fkM = - \sum_{k=1}^\infty \frac{(-1)^kRkt_k}{2}
          \left((\fkL^k)_{>0} - (\fkL^k)_{<0}\right),
  \label{5D-M-formula1}.
\end{gather}
(\ref{5D-ab-eq1bis}) coincides with (\ref{5D-betaLam-eq1}), 
if the building blocks of the Riemann-Hilbert problem 
and the generalized string equations are identified as 
\beq
  Q = (R\Lambda_0)^2,\quad 
  a = R\Lambda,\quad b = - \beta,\quad 
  y = p, \quad Z = e^{-Rz} = - \fkL.  
\label{5D-RH-vs-dToda}
\eeq
Note that the sign factor $(-1)^k$ in (\ref{5D-T=-Tbar}) 
is correlated with the negative sign in the last relation 
of (\ref{5D-RH-vs-dToda}) between $Z$ and $\fkL$. 
Comparing (\ref{5D-M-formula1}) with (\ref{5D-N(z)-Z(y)}), 
one finds that $W(z)$ and $\fkM$ are related as 
\beq
  W(z) + \frac{R}{2}(z-s) + \log y(z) + \frac{V'(z)}{2} 
  = N(z)\sqrt{P(z)} 
  = - \fkM. 
\eeq
These are all that one can derive from the generalized 
string equations; (\ref{5D-betaLam-eq2}) is missing here, 
and has to be added by hand for $a$ and $b$ to be determined. 
Thus the aforementioned ``arbitrariness'' is not resolved 
even in the diagonal subspace (\ref{5D-T=-Tbar}).

\subsection{Deriving generalized string equations}

We now turn to the second subject of this section, 
namely, the origin of the generalized string equations 
in hidden symmetries of a tau function of the Toda hierarchy. 

As observed in our previous work \cite{NT07}, 
the partition function of the 5D theory is connected 
with a tau function $\tau_s(\bsT,\bar{\bsT})$ 
of the Toda hierarchy as 
\begin{multline}
  Z^{\mathrm{5D}}_s(\bst) 
  = \exp\left(\sum_{k=1}^\infty\frac{t_kq^k}{1-q^k}\right)
    q^{-s(s+1)(2s+1)/6}\\
  \mbox{}\times 
    \tau_s(\bsT,\bar{\bsT})|_{T_k=-\bar{T}_k=(-1)^kt_k/2\;(k\ge 1)}. 
\end{multline}
This tau function $\tau_s(\bsT,\bar{\bsT})$ has 
the fermionic expression 
\beq
  \tau_s(\bsT,\bar{\bsT}) 
  = \langle s|\exp\left(\sum_{k=1}^\infty T_kJ_k\right)
    g \exp\left(- \sum_{k=1}^\infty\bar{T}_kJ_{-k}\right)
    |s\rangle, 
\eeq
where $g$ is the $\mathrm{GL}(\infty)$ element 
\beq
  g = q^{W_0/2}G_{-}G_{+}Q^{L_0}G_{-}G_{+}q^{W_0/2}. 
\label{5D-g}
\eeq

It is also pointed out in the same work (loc. cit.) 
that $g$ satisfies the algebraic relations 
\beq
  J_kg = gJ_{-k}, \quad k = 1,2,\cdots. 
\label{5D-Jg=gJ}
\eeq
These relations can be converted to the equations 
\beq
  \left(\frac{\rd}{\rd T_k} + \frac{\rd}{\rd\bar{T}_k}\right)
  \tau_s(\bsT,\bar{\bsT}) = 0, 
  \quad k = 1,2,\cdots,
\eeq
for the tau function, which thereby reduces 
to a function of $\bsT - \bar{\bsT}$: 
\beq
  \tau_s(\bsT,\bar{\bsT}) = \tau^{\mathrm{1D}}_s(\bsT-\bar{\bsT}). 
\eeq
The function $\tau^{\mathrm{1D}}_s(\bsT)$ is a tau function 
of the 1D Toda hierarchy.  

In the language of the Lax operators \cite{NTT95,Takasaki96}, 
(\ref{5D-Jg=gJ}) amount to the relations 
\beq
  L^k = \bar{L}^{-k}, \quad k = 1,2,\cdots,
\eeq
which reduce to the single equation 
\beq
  \fkL := L = \bar{L}^{-1}. 
\label{5D-L=Lbar^{-1}}
\eeq
This implies that $\fkL$ becomes a difference operator 
of the form (\ref{1DToda-Lax-op}), namely 
the Lax operator of the 1D Toda lattice.  
One can thus reconfirm that the foregoing tau function 
is a special solution of the 1D Toda hierarchy.  

Actually, (\ref{5D-Jg=gJ}) are merely a small part  
of a large set of relations satisfied by $g$.  
Those relations are derived from the following 
two types of ``shift symmetries'' \cite{NT07} 
among the elements 
\begin{gather*}
  V^{(k)}_m = q^{-km/2}\sum_{n\in\ZZ}q^{kn}{:}\psi_{m-n}\psi^*_n{:},\\
  \tilde{V}^{(k)}_m = V^{(k)}_m - \frac{q^k}{1-q^k}\delta_{m,0}, 
  \quad k,m \in \ZZ, 
\end{gather*}
of the quantum torus algebra: 
\begin{itemize}
\item First shift symmetry
\beq
  G_{-}G_{+}\tilde{V}^{(k)}_m (G_{-}G_{+})^{-1} 
  = (-1)^k\tilde{V}^{(k)}_{m+k}. 
\eeq
\item Second shift symmetry
\beq
  q^{W_0/2}V^{(k)}_mq^{-W_0/2} = V^{(k-m)}_m 
\eeq
equivalently, 
\beq
  q^{W_0/2}\tilde{V}^{(k)}_mq^{-W_0/2} = \tilde{V}^{(k-m)}_m. 
\eeq
\end{itemize}
Using these symmetries repeatedly, one can rewrite 
$V^{(k)}_mg$ as 
\beqnn
\begin{aligned}
  \tilde{V}^{(k)}_mg 
  &= \tilde{V}^{(k)}_mq^{W_0/2}G_{-}G_{+}Q^{L_0}G_{-}G_{+}q^{W_0/2} \\
  &= q^{W_0/2}\tilde{V}^{(k+m)}_mG_{-}G_{+}Q^{L_0}G_{-}G_{+}q^{W_0/2} \\
  &= (-1)^{k+m}q^{W_0/2}G_{-}G_{+}\tilde{V}^{(k+m)}_{-k}
     Q^{L_0}G_{-}G_{+}e^{W_0/2} \\
  &= (-1)^{k+m}Q^{-k}q^{W_0/2}G_{-}G_{+}Q^{L_0}
     \tilde{V}^{(k+m)}_{-k}G_{-}G_{+}q^{W_0/2} \\
  & = (-1)^{k+m}(-1)^{k+m}Q^{-k}q^{W_0/2}G_{-}G_{+}Q^{L_0}
     G_{-}G_{+}\tilde{V}^{(k+m)}_{-m-2k}q^{W_0/2} \\
  &= Q^{-k}q^{W_0/2}G_{-}G_{+}Q^{L_0}G_{-}G_{+}q^{W_0/2}
     \tilde{V}^{(-k)}_{-m-2k} \\
  &= Q^{-k}g\tilde{V}^{(-k)}_{-m-2k}, 
\end{aligned}
\eeqnn
thus obtains the relations  
\beq
  \tilde{V}^{(k)}_mg = gQ^{-k}\tilde{V}^{(-k)}_{-m-2k}. 
\label{5D-Vg=gV}
\eeq
(\ref{5D-Jg=gJ}) are contained therein as a special case 
where $k = 0$ and $m \ge 1$. 

When $m = 0$ and $k \ge 1$, (\ref{5D-Vg=gV}) become 
the relations 
\beq
  \tilde{V}^{(k)}_0g = gQ^{-k}\tilde{V}^{(-k)}_{-2k}. 
\eeq
According to the general rule \cite{NTT95,Takasaki96,Takasaki10}, 
$\tilde{V}^{(k)}_m$ and $\tilde{V}^{(-k)}_{-2k}$ 
correspond to $q^{-km/2}q^{kM}L^m$ 
and $q^{-km/2}q^{k\bar{M}}\bar{L}^m$, 
and these relations imply the equations 
\beq
  q^{kM} = Q^{-k}q^{-k^2}\bar{L}^{-2k}q^{-k\bar{M}},
  \quad k = 1,2,\cdots, 
\eeq
for the Lax and Orlov-Schulman operators.  
Taking into account the twisted canonical relations 
(\ref{Toda-CCR}), one can easily see that 
these equations reduce to the single equation
\beq
  q^M = (Qq)^{-1}\bar{L}^{-2}q^{-\bar{M}}. 
\label{5D-q^M=...}
\eeq

Thus the special solution of the Toda hierarchy 
determined by (\ref{5D-g}) turns out 
to satisfy (\ref{5D-L=Lbar^{-1}}) and (\ref{5D-q^M=...}). 
These equations may be thought of as generalized 
string equations in the Toda hierarchy. Lastly, 
setting $q = e^{-R\hbar}$, rescaling $M$ and $\bar{M}$ as 
(\ref{rescaling-sTTbar}) and letting $\hbar \to 0$, 
these equations turn into the generalized string equations 
(\ref{5D-GSE}) for the dispersionless Toda hierarchy.

\section{Concluding remarks} 

We have identified the Seiberg-Witten curves 
(\ref{4D-SWcurve}) and (\ref{5D-SWcurve}) 
for the deformed $U(1)$ gauge theories.  
What about the Seiberg-Witten differential 
and prepotential?  This issue is briefly discussed 
in our previous paper \cite{NNT08} 
(though not in a fully correct form, because 
of the incorrect formulation of the Riemann-Hilbert problem). 
We reconsider this issue in the present setup. 

After the idea proposed in the previous paper, 
we consider the critical values 
\beqnn
  \calE^{\mathrm{4D}(0)}_*
  = \calE^{\mathrm{4D}(0)}_{\bst}[\rho^{(0)}_*],\quad 
  \calE^{\mathrm{5D}(0)}_*
  = \calE^{\mathrm{5D}(0)}_{s,\bst}[\rho^{(0)}_*] 
\eeqnn
of the energy functionals as functions of $(s,\bst)$.  
As the following calculations demonstrate, 
they may be thought of as the prepotentials.  

Let us show that the $\bst$-derivatives 
of these functions can be expressed as 
\begin{gather}
  \frac{\rd\calE^{\mathrm{4D}(0)}_*(s,\bst)}{\rd t_k} 
  = \int_{u_0}^{u_1}du\frac{u^{k+1}}{k+1}\rho^{(0)\prime}_*(u),
  \label{4D-E*-derivative1}\\
  \frac{\rd\calE^{\mathrm{5D}(0)}_*(s,\bst)}{\rd t_k} 
  = \int_{u_0}^{u_1}du\frac{e^{-Rku}}{-Rk}\rho^{(0)\prime}_*(u).
  \label{5D-E*-derivative1}
\end{gather}
Since the treatment of the 4D and 5D cases are parallel, 
we illustrate the calculations for the 4D case only. 
Since $\rho^{(0)}_*$ is a solution of (\ref{4D-rho0'-eq1}), 
the identity 
\beqnn
  \mathrm{PP}\int_{u_0}^{u_1}dv 
   g_{\mathrm{4D}}^{(0)\dprime}(|u-v|)\rho^{(0)\prime}_*(v)
  + \frac{V'(u)}{2}  
  = 0 \quad (u_0 \le u \le u_1) 
\eeqnn
holds. Integrated twice with respect to $u$, 
it turns into the identity 
\footnote{The variational equations (\ref{4D-rho0'-eq}) 
and (\ref{5D-rho0'-eq}) can be recovered from 
(\ref{4D-rho0'-eq1}) and (\ref{5D-rho0'-eq1}) 
by differentiating once. Therefore $C_1$ can be 
identified with $\nu/2$.}
\begin{multline}
  \int_{u_0}^{u_1}dv 
  g_{\mathrm{4D}}^{(0)}(|u-v|)\rho^{(0)\prime}_*(v) 
  + \frac{1}{2}\sum_{k=1}^\infty\frac{t_ku^{k+1}}{k+1}
  + C_1u + C_2 = 0 \\
  (u_0 \le u \le u_1),
\label{4D-rho0'-eq2}
\end{multline}
where $C_1$ and $C_2$ are functions of $\bst$ only. 
Differentiating the explicit form 
\begin{multline*}
  \calE^{\mathrm{4D}(0)}_* 
  = \iint_{u_0}^{u_1} dudvg_{\mathrm{4D}}^{(0)}(|u-v|)
     \rho^{(0)\prime}_*(u)\rho^{(0)\prime}_*(v)\\
    + \int_{u_0}^{u_1}du
      \left(\sum_{k=1}^\infty\frac{t_ku^{k+1}}{k+1}
      \right)\rho^{(0)}_*(u) 
\end{multline*}
of the critical value with respect to $t_k$ yields 
\begin{multline*}
  \frac{\calE^{\mathrm{4D}(0)}_*}{\rd t_k}
  = \int_{u_0}^{u_1} du\frac{\rd\rho^{(0)\prime}_*(u)}{\rd t_k}
    \left(
       2 \int_{u_0}^{u_1} dvg_{\mathrm{4D}}^{(0)}(|u-v|)\rho^{(0)\prime}_*(v)
       + \sum_{k=1}^\infty\frac{t_ku^{k+1}}{k+1}
    \right) \\
   + \int_{u_0}^{u_1} du\frac{u^{k+1}}{k+1}\rho^{(0)\prime}_*(u). 
\end{multline*}
We can use (\ref{4D-rho0'-eq2}) to simplify the part 
inside the parenthesis as 
\beqnn
  \frac{\calE^{\mathrm{4D}(0)}_*}{\rd t_k}
  = - 2\int_{u_0}^{u_1} du
        \frac{\rd\rho^{(0)\prime}_*(u)}{\rd t_k}(C_1u + C_2)
    + \int_{u_0}^{u_1} du\frac{u^{k+1}}{k+1}\rho^{(0)\prime}_*(u). 
\eeqnn
On the other hand, since $\rho^{(0)}_*$ satisfies 
(\ref{rho0'-constraint}), the identities 
\beqnn
  \int_{u_0}^{u_1} du\rho^{(0)\prime}_*(u) = -1, \quad 
  \int_{u_0}^{u_1} duu\rho^{(0)\prime}_*(u) = 0 
\eeqnn
hold.  Differentiating them with respect to $t_k$ gives 
\beqnn
  \int_{u_0}^{u_1} du\frac{\rd\rho^{(0)\prime}_*(u)}{\rd t_k} = 0,\quad 
  \int_{u_0}^{u_1} duu\frac{\rd\rho^{(0)\prime}_*(u)}{\rd t_k} = 0. 
\eeqnn
Therefore the integral containing $C_1u + C_2$ vanishes, 
and we obtain (\ref{4D-E*-derivative1}). 

By (\ref{rho0*-ImW}), we can rewrite the right hand side 
of (\ref{4D-E*-derivative1}) and (\ref{4D-E*-derivative1}) 
to contour integrals as 
\begin{gather}
  \frac{\rd\calE^{\mathrm{4D}(0)}_*(s,\bst)}{\rd t_k} 
  = \frac{1}{2\pi i}\oint_C dz\frac{z^{k+1}}{k+1}dW(z), 
  \label{4D-E*-derivative2}\\
  \frac{\rd\calE^{\mathrm{5D}(0)}_*(s,\bst)}{\rd t_k} 
  = \frac{1}{2\pi i}\oint_C dz\frac{e^{-Rkz}}{-Rk}dW(z), 
  \label{5D-E*-derivative2}
\end{gather}
where $C$ is a simple closed curve encircling 
the interval $[u_0,u_1]$ anti-clockwise. 
Note that $dW(z)$ is a single-valued differential, 
so that the contour integrals are meaningful 
in the usual sense.  Unfortunately, $W(z)$ is multi-valued 
on the cut plane, so that one {\it cannot} do 
integration by part to replace the differentials as 
\beqnn
  \frac{z^{k+1}}{k+1}dW(z) \to - z^kW(z)dz, \quad 
  \frac{e^{-Rkz}}{-Rk}dW(z) \to - e^{-Rkz}W(z)dz. 
\eeqnn
Apart from this problem, $W(z)dz$ may be thought of 
as a candidate of the Seiberg-Witten differential.  
Actually, one can add the exact form $dV(z)/2$ 
to replace it by 
\beqnn
  \frac{dS(z)}{2} = W(z)dz + \frac{dV(z)}{2}, 
\eeqnn
where $S(z)$ is the $S$-function (\ref{S-function}).  
Thus we eventually reach the same conclusion as 
Marshakov and Nekrasov \cite{MN06,Marshakov07} 
that the total differential of the $S$ function 
gives the Seiberg-Witten differential.  
The numerical factor $2$ is correlated with the denominator 
of (\ref{4D-T=-Tbar}) and (\ref{5D-T=-Tbar}); 
if $T_k$'s (or $\bar{T}_k$'s) are used in place of $t_k$'s, 
this numerical factor disappears.

\subsection*{Acknowledgements}

This work is partly supported by JSPS Grants-in-Aid 
for Scientific Research No. 19104002, No. 21540218 
and No. 22540186 from the Japan Society for the Promotion of Science.


\begin{thebibliography}{99}

\bibitem{NNT08}
T. Nakatsu, Y. Noma and K. Takasaki,
Extended 5d Seiberg-Witten theory and melting crystal, 
Nucl. Phys. {\bf B808} [FS] (2009), 411--440. 

\bibitem{Losev-etal03} 
A. Losev, A. Marshakov and N. Nekrasov, 
Small instantons, little strings and free fermions, 
in: M. Shifman, A. Vainstein and J. Wheater (eds.), 
{\it From fields to strings: circumnavigating theoretical physics\/}
(World Scientific, 2005), pp. 581--621, 

\bibitem{OR03}
A. Okounkov and N. Reshetikhin,  
Correlation function of Schur Process with application 
to local geometry of a random 3-Dimensional Young diagram, 
J. Amer. Math. Soc. {\bf 16}, (2003), 581--603. 

\bibitem{Nekrasov02}
N. Nekrasov,
Seiberg-Witten Prepotential from Instanton Counting, 
Adv. Theor. Math. Phys. {\bf 7}  (2004), 831--864. 

\bibitem{ORV03}
A. Okounkov, N. Reshetikhin and C. Vafa,  
Quantum Calabi-Yau and classical crystals, 
in: P. Etingof, V. Retakh and I.M. Singer (eds.), 
{\it The unity of mathematics\/}, 
Progr. Math.  vol. 244 (Birkh\"auser, 2006), 
pp. 597--618. 

\bibitem{MNTT04a}
T. Maeda, T. Nakatsu, K. Takasaki and T. Tamakoshi, 
Five-dimensional supersymmetric Yang-Mills theories 
and random plane partitions, 
JHEP {\bf 0503}  (2005), 056. 

\bibitem{MNTT04b}
T. Maeda, T. Nakatsu, K. Takasaki and T. Tamakoshi, 
Free fermion and Seiberg-Witten differential in 
random plane partitions, 
Nucl. Phys. {\bf B715} (2005), 275--303. 

\bibitem{NT07}
T. Nakatsu and K. Takasaki, 
Melting crystal, quantum torus and Toda hierarchy, 
Commun. Math. Phys. {\bf 285} (2009), 445--468. 

\bibitem{NT08}
T. Nakatsu and K. Takasaki, 
Integrable structure of melting crystal model with external potentials, 
Advanced Studies in Pure Mathematics vol. 59 
(Mathematical Society of Japan, 2010), pp. 201--223. 

\bibitem{SW94}
N. Seiberg and E. Witten,
Electric-magnetic duality, monopole condensation, 
and confinement in $N=2$ supersymmetric Yang-Mills theory, 
Nucl. Phys. {\bf B426} (1994) 19--52; 
Erratum,  ibid.\ \textbf{B430} (1994) 485; 
Monopoles, duality and chiral symmetry breaking 
in $N=2$ Supersymmetric QCD, 
ibid. {\bf B431} (1994), 494--550. 

\bibitem{NO03}
N. Nekrasov and A. Okounkov,
Seiberg-Witten theory and random partitions, 
in: P. Etingof, V. Retakh and I.M. Singer (eds.), 
{\it The unity of mathematics\/}, 
Progr. Math.  {\bf 244}, Birkh\"auser, 2006, 
pp. 525--296. 

\bibitem{MN06}
A. Marshakov and N. Nekrasov,  
Extended Seiberg-Witten theory and integrable hierarchy, 
JHEP {\bf 0701}  (2007), 104. 

\bibitem{Marshakov07}
A. Marshakov, 
On microscopic origin of integrability in Seiberg-Witten theory, 
Theor. Math. Phys. {\bf 154} (2008), 362--384. 

\bibitem{Maeda-Nakatsu06}
T. Maeda and T. Nakatsu, 
Amoeba and instantons, 
Int. J. Mod. Phys. {\bf A22} (2007), 937--984. 

\bibitem{TT91}
K. Takasaki and T. Takebe, 
$\mathrm{SDiff(2)}$ Toda equation --- hierarchy, 
tau function and symmetries, 
Lett. Math. Phys. {\bf 23} (1991), 491--503. 

\bibitem{TT93}
K. Takasaki and T. Takebe, 
Quasi-classical limit of Toda hierarchy and W-infinity symmetries, 
Lett. Math. Phys. {\bf 28} (1993), 165-176.  

\bibitem{UT84}
K. Ueno and K. Takasaki, 
Toda lattice hierarchy, 
Adv. Stud. Pure Math. {\bf 4} (1984), 1--95. 

\bibitem{TT95}
K. Takasaki, and T. Takebe, 
Integrable hierarchies and dispersionless limit, 
Rev. Math.  Phys. {\bf 7} (1995), 743--808. 

\bibitem{EK94}
T. Eguchi and H. Kanno, 
Toda lattice hierarchy and the topological description 
of $c = 1$ string theory, 
Phys. Lett. {\bf B331} (1994), 330--334.  

\bibitem{Takasaki94}
K. Takasaki, 
Dispersionless Toda hierarchy and two-dmensional string theory, 
Commun. Math. Phys. {\bf 170} (1995), 101--116. 

\bibitem{NTT95}
T. Nakatsu, K. Takasaki and S. Tsujimaru, 
Quantum and classical aspects of deformed $c = 1$ strings, 
Nucl. Phys. {\bf B443} (1995), 155--197.  

\bibitem{KO95}
H. Kanno and Y. Ohta,
Topological strings with scaling violation and
Toda lattice hierarchy, 
Nucl. Phys. {\bf B442} (1995), 179--204. 

\bibitem{Takasaki96}
K. Takasaki, 
Toda lattice hierarchy and generalized string equations, 
Commun. Math. Phys. {\bf 181} (1996), 131--156.  

\bibitem{WZ00}
P.~B. Wiegmann and A. Zabrodin,
Conformal maps and integrable hierarchies,
Comm. Math. Phys. {\bf 213} (2000), 523--538. 

\bibitem{MWWZ00}
M. Mineev-Weinstein, P.~B. Wiegmann and A. Zabrodin, 
Integrable structure of interface dynamics, 
Phys. Rev. Lett. {\bf 84} (2000), 5106--5109. 

\bibitem{Zabrodin01}
A. Zabrodin,
Dispersionless limit of Hirota equations
in some problems of complex analysis,
Theor. Math. Phys. {\bf 12} (2001), 1511--1525. 

\bibitem{Teo09}
L.-P. Teo,
Conformal mappings and dispersionless Toda hierarchy II:
General string equations,
Commun. Math. Phys. {\bf 297} (2010), 447--474. 

\bibitem{Takasaki10}
K. Takasaki, 
Generalized string equations for double Hurwitz numbers, 
arXiv:1012.5554 [math-ph]. 

\bibitem{Sagan-book}
B.E. Sagan, 
{\it The Symmetric groups: Representations, 
combinatorial algorithms, and symmetric functions\/}, 
Springer-Verlag, 2001.  

\bibitem{Macdonald-book} 
I.~G. Macdonald,
{\it Symmetric functions and Hall polynomials\/}, 
Oxford University Press, 1995.

\bibitem{Nishizawa95}
M. Nishizawa,
On a $q$-Analogue of the Multiple Gamma Functions,
Lett. Math. Phys. {\bf 37} (1996), 201--209. 

\bibitem{Flaschka74}
H. Flaschka, 
The Toda lattice I, Existence of integrals, 
Phys. Rev. {\bf B9} (1974), 1924--1925; 
The Toda lattice II, Inverse scattering solution, 
Prog. Theor. Phys. {\bf 51} (1974), 703--716. 

\bibitem{Adler79}
M. Adler, 
On a trace functional for formal pseudo-differential operators 
and the symplectic structure of the Korteweg-deVries type equations, 
Invent. Math. {\bf 50} (1979), 219--248.  

\bibitem{LS77}
B.~F. Logan and L.~A. Shepp, 
A variational problem for random Young tableau, 
Adv. Math. {\bf 26} (1977), 206--202. 

\bibitem{VK77}
A.~M. Vershik and S.~V. Kerov, 
Asymptotics of the Plancherel measure of the symmetric group 
and the limiting form of Young tableaux, 
Soviet Math. Dokl. {\bf 18} (1977), 527--531. 

\bibitem{EY94}
T. Eguchi and S.-K. Yang,
The topological $CP^1$ model and the large $N$ matrix integral,
Mod. Phys. Lett. {\bf A9} (1994), 2893--2902.

\bibitem{MAM07}
L. Mart\'inez Alonso and E. Medina, 
Semiclassical expansions in the Toda hierarchy 
and the Hermitian matrix model, 
J. Phys. A: Math. Theo. {\bf 40} (2007), 14223--14241. 

\end{thebibliography}
\end{document}